\documentclass[sigconf, nonacm, appendixcol=same]{acmart}

\pdfoutput=1

\usepackage{xspace}
\usepackage{enumerate}
\usepackage{algorithm} 
\usepackage[noend]{algpseudocode} 
\usepackage{ifsym}

\usepackage{caption}
\usepackage{subcaption}

\usepackage{comment}

\usepackage{lineno}

\usepackage{xcolor}

\usepackage{tikz}
\usepackage{collcell}
\usetikzlibrary{calc}
\usetikzlibrary{arrows}
\usetikzlibrary{positioning}
\usepackage{graphicx}
\usepackage{subcaption}
\usepackage{float}
\usepackage{pgf-pie}
\usepackage{pgfplots}
\pgfplotsset{compat=1.16}
\usepackage{lipsum}

\usepackage[prefix=s]{xcolor-solarized}

\usepackage{apxproof}

\theoremstyle{plain}
\newtheoremrep{theorem}{Theorem}[section]
\newtheoremrep{lemma}[theorem]{Lemma}
\newtheoremrep{claim}[theorem]{Claim}
\newtheoremrep{corollary}[theorem]{Corollary}
\newtheoremrep{proposition}[theorem]{Proposition}
\newtheoremrep{conjecture}[theorem]{Conjecture}

\usepackage{listings}
\lstdefinelanguage{pgkeysLang}
{
  morekeywords={
    MATCH,
    WITH,
    WHERE,
    COUNT,
    FOR,
    WITHIN,
    EXCLUSIVE,
    MANDATORY,
    SINGLETON,
    IDENTIFIER,
    MATCH,
    WHERE,
    FROM,
    NOT,
    OR,
    AND,
    EXISTS,
    RETURN,
    IN,
    IS,
    NULL,
    SELECT,
    LIMIT,
    HAVING,
    VALUES,
    GROUP,
    BY,
    AS,
    UNWIND,
    OPTIONAL,
    hasKey,
    FunctionalProperty,
    context,
    test,
    unique,
    key,
    selector,
    field,
    allInstances,
    isUnique,
    Tuple,
    class,
    extent,
    relationship,
    inverse,
    attribute,
    source,
    target,
    hops,
    filter,
    edges,
    label,
    timeRange,
    laterThan
  },
  sensitive=false, 
  morecomment=[l]{//}, 
  morestring=[b]{\'} 
}

\definecolor{eclipseBlue}{RGB}{42,0.0,255}
\definecolor{eclipseGreen}{RGB}{63,127,95}
\definecolor{eclipsePurple}{RGB}{127,0,85}

\newcommand{\mcomment}[2]{{\color{blue}\textbf{(#1)}}\footnote{\textbf{#1:} #2}}

\newcommand{\wim}[1]{\mcomment{Wim}{#1}}

\newcommand{\red}[1]{{\color{red}#1}}
\definecolor{RoyalBlue}{RGB}{65,105,225}
\newcommand{\newstuff}[1]{{\color{black}#1}}

\newcommand{\nat}{\ensuremath{\mathbb{N}}}

\newcommand{\limit}{\textsf{LIMIT}\xspace}
\newcommand{\orderby}{\textsf{ORDER BY}\xspace}

\newcommand{\crpq}{\text{CRPQ}\xspace}
\newcommand{\crpqs}{\text{CRPQ}s\xspace}
\newcommand{\crpqf}{\text{CRPQ}$_f$\xspace}
\newcommand{\crpqfs}{\text{CRPQ}$_f$s\xspace}

\newcommand{\true}{\ensuremath{\mathtt{true}}\xspace}
\newcommand{\false}{\ensuremath{\mathtt{false}}\xspace}

\newcommand{\coNP}{\ensuremath{\mathsf{coNP}}\xspace}
\newcommand{\conp}{\coNP}
\newcommand{\NP}{\ensuremath{\mathsf{NP}}\xspace}
\newcommand{\np}{\NP}

\renewcommand{\P}{\ensuremath{\mathsf{P}}\xspace}
\newcommand{\ptime}{\P}

\newcommand{\lab}{\ensuremath{\mathsf{Lab}}\xspace}

\newcommand{\prop}{\ensuremath{\mathsf{Prop}}\xspace}

\newcommand{\bG}{\ensuremath{\mathbf{G}}\xspace}

\newcommand{\nodeid}{\ensuremath{\mathrm{NID}}\xspace}
\newcommand{\edgeid}{\ensuremath{\mathrm{EID}}\xspace}
\newcommand{\const}{\ensuremath{\mathsf{Const}}\xspace}
\newcommand{\rel}{\ensuremath{\mathsf{Rel}}\xspace}
\newcommand{\edge}{\mathrm{Edge}\xspace}
\newcommand{\node}{\mathrm{Node}\xspace}

\newcommand{\adom}{\mathsf{Adom}\xspace}

\newcommand{\dom}{\mathsf{Dom}\xspace}
\newcommand{\var}{\mathsf{Var}\xspace}
\newcommand{\fvar}{\mathsf{FVar}\xspace}
\newcommand{\ftvar}{\mathsf{FTVar}\xspace}
\newcommand{\qvar}{\mathsf{QVar}\xspace}
\newcommand{\tvar}{\mathsf{TVar}\xspace}
\newcommand{\tw}{\mathsf{tw}\xspace}

\newcommand{\ssz}{\mathsf{ss}\xspace}
\newcommand{\fctw}{\mathsf{fc\textsf{-}tw}\xspace}
\newcommand{\cR}{\ensuremath{\mathcal{R}}\xspace}

\newcommand{\OMIT}[1]{}

\newcommand{\fullversion}[2]{#1}
\begin{document}
\begin{abstract}
Threshold queries are an important class of queries that only require computing or counting answers up to a specified threshold value. To the best of our knowledge, threshold queries have been largely disregarded in the research literature, which is surprising considering how common they are in practice. In this paper, we present a deep theoretical analysis of threshold query evaluation and show that thresholds can be used to significantly improve the asymptotic bounds of state-of-the-art query evaluation algorithms.
We also empirically show that threshold queries are significant in practice. In surprising contrast to conventional wisdom, we found important scenarios in real-world data sets in which users are interested in computing the results of queries up to a certain threshold, independent of a ranking function that orders the query results by importance. 
\end{abstract}

\title{Threshold Queries in Theory and in the Wild}

\settopmatter{authorsperrow=5}

\author{Angela Bonifati}
\affiliation{\small Lyon 1 Univ.}

\author{\mbox{Stefania Dumbrava}}
\affiliation{\small ENSIIE}

\author{George Fletcher}
\affiliation{\small TU Eindhoven}

\author{Jan Hidders}
\affiliation{\small Univ. of London, Birkbeck} 

\author{Matthias Hofer}
\affiliation{\small Univ. of Bayreuth}

\author{Wim Martens}
\affiliation{\small Univ. of Bayreuth}

\author{Filip Murlak}
\affiliation{\small Univ. of Warsaw}

\author{Joshua Shinavier}
\affiliation{\small Uber Inc.}

\author{Sławek Staworko\rlap{$^\dagger$}}\thanks{$\dagger$Contact author: \url{slawomir.staworko@univ-lille.fr}}
\affiliation{\small Univ. of Lille}

\author{\mbox{Dominik Tomaszuk}}
\affiliation{\small Univ. of Bialystok}

\renewcommand{\shortauthors}{A.\ Bonifati et al.}


\maketitle

\newcommand\vldbdoi{XX.XX/XXX.XX}
\newcommand\vldbpages{XXX-XXX}
\newcommand\vldbvolume{14}
\newcommand\vldbissue{1}
\newcommand\vldbyear{2020}
\newcommand\vldbauthors{\authors}
\newcommand\vldbtitle{\shorttitle} 
\newcommand\vldbavailabilityurl{http://vldb.org/pvldb/format_vol14.html}

\begingroup\small\noindent\raggedright\textbf{PVLDB Reference Format:}\\
\vldbauthors. \vldbtitle. PVLDB, \vldbvolume(\vldbissue): \vldbpages, \vldbyear.\\
\href{https://doi.org/\vldbdoi}{doi:\vldbdoi}
\endgroup
\begingroup
\renewcommand\thefootnote{}\footnote{\noindent
This work is licensed under the Creative Commons BY-NC-ND 4.0 International License. Visit \url{https://creativecommons.org/licenses/by-nc-nd/4.0/} to view a copy of this license. For any use beyond those covered by this license, obtain permission by emailing \href{mailto:info@vldb.org}{info@vldb.org}. Copyright is held by the owner/author(s). Publication rights licensed to the VLDB Endowment. \\
\raggedright Proceedings of the VLDB Endowment, Vol. \vldbvolume, No. \vldbissue\ %
ISSN 2150-8097. \\
\href{https://doi.org/\vldbdoi}{doi:\vldbdoi} \\
}\addtocounter{footnote}{-1}\endgroup

\ifdefempty{\vldbavailabilityurl}{}{
\vspace{.3cm}
\begingroup\small\noindent\raggedright\textbf{PVLDB Availability Tag:}\\
The source code of this research paper has been made publicly available at \url{\vldbavailabilityurl}.
\endgroup
}



\section{Introduction}
\label{sec:introduction}


Queries encountered in a wide range of data management applications,
such as data interaction, data visualization, data exploration, data curation and data monitoring~\cite{BattleS21,IdreosPC15,IlyasC19},
often require computing or counting answers only up to a given threshold value. 
We call such queries \emph{threshold queries}.



\smallskip

\noindent {\em Threshold queries for data exploration. }
Querying voluminous rich data sources, such as Wikidata~\cite{vrandecic2012},  may return more results than are needed during exploratory analytics. Thus, users may specify a threshold
on the number of answers they want to see. Consider the following query which lists up to a threshold (expressed with the \texttt{LIMIT} clause in SPARQL) businesses, total assets, cities, and countries of headquarters locations in the Wikidata dataset. Without \texttt{LIMIT}, this query would return 2,684,138 objects, amounting to about 413MB in size, which is too large for human consumption. 
\begin{lstlisting}
SELECT ?business ?assets ?city ?country {
  ?business <total_assets> ?assets .
  ?business <headquarters_location> ?city .
  ?city <country> ?country .
} LIMIT 10
\end{lstlisting}
%
%
%
Note that here the user is interested in  unranked output (i.e., there is no \texttt{ORDER BY} clause), which is typical for data exploration \cite{BattleS21}.
As we will see later in a deep-dive into real query logs (Section~\ref{sec:use-cases}), such queries on Wikidata are very common in practice. 




\smallskip

\noindent {\em Threshold queries for data curation.} 
Threshold queries are also useful in detecting violations of database constraints and identifying data items requiring curation actions.
As an example, consider the following threshold query using  \emph{grouping} and aggregation in
the Nobel Prize database, requiring that every Nobel prize has at most three laureates~\cite{RaMA17}.
\begin{lstlisting}[language=SQL]
 SELECT P.ID FROM NobelPrize P, Laureate L
 WHERE P.ID = L.Prize_ID
 GROUP BY P.ID
 HAVING COUNT(*) > 3.
\end{lstlisting}
\noindent 
{\bf Differences with other query answering paradigms.} Although threshold queries might look similar in spirit to top-$k$ queries~\cite{ilyas2004supporting,natsev2001supporting,finger2009robust,mamoulis2007efficient},
they are inherently different because they do not assume that the results are ranked.  They are also different from counting queries~\cite{DurandM15, Pichler_Skritek_2013}, since these aim at computing an exact value, rather than only desiring exact values up to a given threshold. 
Therefore, prior problems are either more specific or have different objectives. We examine these differences in depth in  Section~\ref{sec:related-work}.

\smallskip

\noindent {\bf Our contributions.} We are motivated by the following question: Can we exploit the fact that we only need to count or compute the answers of a query \emph{up to a threshold}?
In this paper, we answer this question positively. 
The starting point for our work is the observation that evaluating some queries with a threshold $k$ requires storing not more than $k+1$ intermediate results \cite{Carey1997,Kim2018}. 
We show that this idea can be fully integrated with state-of-the-art complex join algorithms, leading to significant savings in the size of the intermediate results as typically computed in query processing, 
leading for some queries to improvements from $\tilde{O}(n^f)$ to $\tilde{O}(n\cdot k)$, where $f$ is the \emph{free-connex treewidth} of the query and $k$ is the value of the threshold. Here, the free-connex treewith of a query measures its treewidth in connection with its output variables. It can be large even if the query is acyclic.

In detail, the key contributions of our paper are as follows:
\begin{enumerate}
\item New results explaining the interplay between different structural properties of conjunctive queries (i.e., select-project-join queries) used in sophisticated evaluation algorithms \newstuff{(Lemma~\ref{lem:interplay})}, and the consequences for threshold query processing \newstuff{(Theorem~\ref{thm:poly})}.  
\item New evaluation algorithms for threshold queries (for computing answers, counting answers, and sampling answers) with improved asymptotic guarantees \newstuff{(Theorem~\ref{thm:pseudopoly})}. 
\item A comprehensive empirical study of threshold queries found in the wild, which highlights their characteristics and shows that they are quite common in important practical scenarios.
\item \newstuff{An experimental evaluation of a proof-of-concept SQL implementation of our algorithm against the current query optimizer in PostgreSQL, showing speedups of several orders of magnitude.}

\end{enumerate}
Our work provides the first in-depth theoretical treatment of threshold queries. In addition, it also shows that these queries are important in practical settings by means of an in-depth empirical study.

The paper is organized as follows. Section~\ref{sec:preliminaries} contains basic definitions. 
In Section~\ref{sec:exploiting-thresholds}, we identify the problems of
interest, explain the complexity-theoretic limits, and explain some of our main ideas in an example. Section~\ref{sec:technical-results}
presents the 
algorithms for the problems of interest. 
In Section~\ref{sec:use-cases}, we present experiments and findings on 
threshold queries that can be found in practice.
In Section~\ref{sec:related-work}, we discuss related
work. Finally, in Section~\ref{sec:conclusions} we summarize our findings and outline further
research directions. 
\fullversion{
Detailed proofs and additional material can found in the appendix.
}{
Because of space limitation, detailed proofs as well as additional material can be found in an extended technical report~\cite{fullArxiv}.
}

\section{Preliminaries}\label{sec:preliminaries}


Our results apply in both a relational database setting and a graph database setting. 
We first focus on the relational database setting, for which we loosely follow the preliminaries as presented in \cite{ABLMP21}.
We assume that we have countably infinite and disjoint sets \rel of \emph{relation names} and \const of \emph{values}. Furthermore, when $(a_1,\ldots,a_k)$ is a Cartesian tuple, we may abbreviate it as $\bar a$. A $k$-ary \emph{database tuple} (henceforth abbreviated as \emph{tuple}) is an element of $\const^k$ for some $k \in \nat$. A \emph{relation} is a finite set $S$ of tuples of the same arity. We denote the set of all such relations by $\cR$.  A \emph{database} is a partial function $D : \rel \to \cR$ such that $\dom(D)$ is finite.
If $D$ is a database and $R \in \rel$, we write $R(a_1,\ldots,a_k) \in D$ to denote that $(a_1,\ldots,a_k) \in D(R)$. By $\adom(D)$ we denote the set of constants appearing in tuples of $D$, also known as the \emph{active domain} of $D$.

For defining conjunctive queries, we assume a countably infinite set $\var$ of
variables, disjoint from 
\rel and \const. An \emph{atom} is an expression of the form
$R(u_1,\ldots,u_k)$, where $u_i \in \var \cup \const$ for each $i \in \{1,\ldots,k\}$. 
A \emph{conjunctive query (CQ)} is an expression $q$ of the form
\[\exists \bar y \; A_1(\bar u_1)\land \ldots \land A_n(\bar u_n) \ , \] 
where $\bar y = (y_1,\ldots,y_m)$ consists of  \emph{existentially quantified variables} and each $A_i(\bar{u}_i)$, with
$i \in \{1,\ldots,n\}$ is an atom. Each variable that appears in $\bar y$ should also appear in $\bar u_1,\ldots,\bar u_n$. On the other hand, $\bar u_1,\ldots,\bar u_n$ can contain variables not present in $\bar y$. For a tuple $\bar x$, we write $q(\bar x)$ to emphasize that $q$ is a CQ such that all variables in $\bar u_1,\ldots,\bar u_n$ appear in either
$\bar x$ or $\bar y$. Unless we say otherwise, we assume that the variables in $\bar u_1,\ldots,\bar u_n$ are \emph{precisely} the variables in $\bar x$ and $\bar y$.
The \emph{arity} of $q(\bar x)$ is defined as the arity of the tuple $\bar x$. We denote by $\var(q)$ the set of all variables appearing in $q$ and by $\fvar(q)$ the set 
of so-called \emph{free variables} of $q$, which are the variables in $\var(q)$ that are not existentially quantified. A \emph{full CQ} is a CQ without existentially quantified variables. 



\paragraph{Query Answers and Relational Algebra} 
\newstuff{We consider queries under \emph{set semantics}, i.e., each answer occurs at most once in the result.} 
A \emph{binding of $X \subseteq \var$} is a function $\eta : X \to \const$.
We say that bindings $\eta$ and $\eta'$ are \emph{compatible} if $\eta(x) = \eta'(x)$ for all $x \in \dom(\eta) \cap \dom(\eta')$.
For compatible bindings $\eta_1$ and $\eta_2$, the \emph{join of $\eta_1$ and $\eta_2$}, is the binding $\eta_1 \bowtie \eta_2$ such that $\dom(\eta_1 \bowtie \eta_2) = \dom(\eta_1) \cup \dom(\eta_2)$ and 
$\big(\eta_1 \bowtie \eta_2\big)(x) = \eta_i(x)$ for all  $x \in \dom(\eta_i)$ and $i\in\{1, 2\}$. If $P_1$ and $P_2$ are sets of bindings, then the \emph{join of $P_1$ and $P_2$} is  $P_1 \bowtie P_2 = \big\{ \eta_1 \bowtie \eta_2 \bigm| \eta_1 \in P_1 \text{ and } \eta_2 \in P_2 \text{ are compatible}\big\}$. For $X \subseteq \dom(\eta)$, the \emph{projection of $\eta$ on $X$}, written as  $\pi_X(\eta)$, is the binding $\eta'$ with $\dom(\eta') = X \cap \dom(\eta)$ and $\eta'(x) = \eta(x)$, for every $x \in \dom(\eta')$. For a set $P$ of bindings, the \emph{projection of $P$ on $X$} is $\pi_X(P) = \big\{\pi_X(\eta)\bigm| \eta \in P\big\}$.

A \emph{match for $q$ in $D$} is a binding of  $\var(q)$ such that $A_i(\eta(\bar u_i)) \in D$ for every $i \in \{1,\ldots,n\}$.\footnote{Notice that here we also denote by $\eta$ the extension of $\eta$ that is the identity on $\const$, and its extension thereof to tuples of variables and constants.} The set of \emph{answers to $q$ on $D$} is $q(D) = \big\{ \pi_{\fvar(q)}(\eta) \bigm| \eta$ is a match for $q$ in $D \big\}$. 
We define answers as functions instead of database tuples, as this simplifies the presentation and as reasoning about their underlying domains is useful in further sections, when dealing with query decompositions. 

\paragraph{Threshold Queries}
A \emph{threshold query (TQ)} is an expression $t$ of the form
\[q(\bar x) \land \exists^{a,b} \bar y\; p(\bar x, \bar y)\]
where, from left to right, $q(\bar x)$ is a CQ, $a\in \mathbb{N}$, $b \in \mathbb{N} \cup \{\infty\}$,  and $p(\bar x, \bar y)$ is a CQ in which we do not require that every variable in $\bar x$ appears in one of its atoms. \newstuff{Notice that a TQ only has a single counting quantifier $\exists^{a,b}$, although further ordinary existential quantifiers may occur inside $q$ and $p$.}
We use $\exists^{\geq a}$ and $\exists^{\leq b}$ as shorthands for $\exists^{a,\infty}$ and $\exists^{0,b}$, respectively, and  the corresponding threshold queries will be called \emph{at-least} and \emph{at-most} queries.
Similar to CQs, we usually denote the entire query as $t(\bar x)$ or even as $t$ when $\bar x$ is clear from the context. When representing $t(\bar x)$ for decision problems, we assume that the numbers $a$ and $b$ are given in binary.

As an example, recall the Nobel Prize threshold query in Section~\ref{sec:introduction}, and suppose that the schema is  $\mathit{NobelPrize}(\underline{\mathit{id}},\mathit{year},\mathit{category})$ and $\mathit{Laureate}(\underline{\mathit{nid}},\underline{\mathit{name}},\mathit{country})$ with the foreign key constraint $\mathit{Laureate}[\mathit{nid}]\subseteq\mathit{NobelPrize}[\mathit{id}]$. This threshold query can be formalized as follows.  
\[
t(x) = 
\exists x_1, x_2.\ \mathit{NobelPrize}(x,x_1,x_2)\land \exists^{{}\geq 4} y.\ \exists z.\ Laureate(x,y,z).
\]

The set of answers of $t$ on $D$, written $t(D)$, is the set of answers $\eta$ of $q(\bar x)$ on $D$ that have between $a$ and $b$ compatible answers of $p(\bar x, \bar y)$. Formally, $\eta \in t(D)$ iff $\eta \in q(D)$ and $a \leq |p(D,\eta)| \leq b$, where $p(D,\eta) = \big\{\eta' \in p(D) \bigm| \eta'$ is \emph{compatible} with $\eta\big\}$.

If $t$ is a threshold query of the form above, we call $\bar x$ the \emph{free variables} (or \emph{answer variables}) of the query $t$ and we write $\fvar(t)$ for the set of these variables. We call $\bar y$ the \emph{tally variables} of the query $t$ and we write $\tvar(t)$ for the set of these variables. 
For the ease of presentation we shall assume that the sets of existentially quantified variables in $q$ and $p$ are disjoint;
we shall write  $\qvar(t)$ for the union of these sets. Thus, $\var(t)$ is the union of three disjoints sets: $\fvar(t)$, $\tvar(t)$, and $\qvar(t)$.



\newstuff{A threshold query $t$ can be intuitively expressed as a SQL query defining $q(\bar x)$ such that in the WHERE clause we additionally check if the number of tuples returned by a correlated SELECT-FROM-WHERE subquery defining $p(\bar x, \bar y)$ is at least $a$ and at most $b$.}

\OMIT{
\paragraph{Threshold queries in SQL}
\newstuff{We illustrate how a threshold $t =
q(\bar x) \land \exists^{a,b} \bar y\; p(\bar x, \bar y)$ can be written in SQL.
Let $q$ be
\[\exists \bar z \; A_1(\bar u_1)\land \ldots \land A_n(\bar u_n) \land \theta_q \ , \] 
and $p$ be
\[\exists \bar w \; B_1(\bar u_1)\land \ldots \land B_m(\bar u_n) \land \theta_p \  \] 
where $\theta_q$ and $\theta_p$ are the join conditions for $q$ and $p$, respectively.  Then an equivalent SQL expression for $t$ can be given as}
\begin{verbatim}
SELECT X
FROM A1,..., An
WHERE theta_q AND EXISTS (SELECT X
                          FROM B1,...,Bm
                          WHERE theta_p AND psi
                          GROUP BY X
                          HAVING COUNT(Y) >= a
                                 AND COUNT(Y) <= b )
\end{verbatim}
\newstuff{where $\psi$ enforces equality on $\bar x$.} \wim{Don't we need a DISTINCT somewhere b/c of set semantics? Like: COUNT(DISTINCT Y)? Also, if we present this we need to clarify the $\theta$s, because CQs express these by identifying variables.}
}
\paragraph{Graph Databases}
For the purposes of this paper, graph databases can be abstracted as relational databases with unary and binary relations. That is, a graph database can be seen as a database $G$ with a unary relation $\node$ and binary relations $A, B, \dots$ where
\begin{itemize}
\item $\node(a) \in G$ if $a$ is a node of the graph database and 
\item $A(a_1,a_2) \in G$  if $(a_1,a_2)$ is an edge with label $A$ in the graph database.
\end{itemize}

An important feature that distinguishes graph database queries from relational database queries are \emph{regular path queries (RPQs)}  \cite{2018Bonifati}. In a nutshell, a regular path query is an expression of the form $r(x,y)$, where $r$ is a regular expression over the set of edge labels in the graph database. When evaluated over a graph database $G$, the query returns all node pairs $(u,v)$ such that there exists a path from $u$ to $v$ that is labeled with some word in the language of $r$. 

All results in the paper extend naturally to RPQs and therefore to so-called conjunctive regular path queries (CRPQs) \cite{2018Bonifati}. This means that we can generalize CQs to CRPQs, which are defined exactly the same as CQs, but we additionally allow RPQs at the level of atoms. Generalizing threshold queries is analogous. If we want to evaluate a threshold query with RPQs, we can pre-evaluate all RPQs in the query and treat their result as a \emph{materialized view}. We can then treat the query as an ordinary threshold query in which each RPQ becomes an atom, which is evaluated over the corresponding materialized view. 

\section{Exploiting thresholds at a glance}
\label{sec:exploiting-thresholds}

Query evaluation is arguably the most fundamental problem in databases and comes in many variants, such as:
\begin{enumerate}[(E1)]
\item Boolean evaluation, i.e.,  testing existence of an answer;
\item returning a given number of answers;
\item counting the total number of answers;
\item sampling answers with uniform probability; and
\item enumerating the answers with small delay.
\end{enumerate}
An important reason why all these variants are considered is that the set of answers to a query can be very large and one is not always interested in the set of all answers. 

The computational cost of these variants tends to increase as we go down in the list,
but already for CQs even the simplest problem (E1) is intractable \cite[Chapter 15]{ABLMP21}.
Triggered by Yannakakis' seminal result on efficient evaluation of \emph{acyclic} CQs \cite{Yannakakis81}, the literature developed a deep understanding that teaches us that, essentially, \emph{low tree-width} is not only helpful but even \emph{necessary} for polynomial-time Boolean evaluation of CQs \cite{GroheSS-stoc01}. Intuitively, the tree-width of a CQ measures the likeness of its graph structure to a tree. In essence, this graph structure is obtained by taking the queries' variables as nodes and connecting variables with an edge iff they appear in a common atom.
Queries with low tree-width are tree-like and queries with high tree-width are highly cyclic. 

\begin{example} \label{ex:boolean}
Consider the following variant of the first query from the introduction: 
\[q(x, y, z) \gets \mathit{Assets}(x,y), \mathit{Subsidiary}(w,x), \mathit{Shareholder}(w,z)\,.\] 
For the purpose of Boolean evaluation we can rewrite $q$ as
\begin{align*}
q'() & \leftarrow   \mathit{Assets}(x,y), U(x)\,; \\
U(x) & \leftarrow  \mathit{Subsidiary}(w,x), V(w)\,; \\ 
V(w) & \leftarrow  \mathit{Shareholder}(w,z)
\end{align*}
using views $U$ and $V$.
It is clear that one can materialize the views and answer $q'$ in time $O\big (n \cdot \log n\big)$ over databases of size $n$.
The tree-width of $q$ manifests itself as the number of variables used in the definitions of the views. For CQs of tree-width $d$, views in the optimal rewriting will use up to $d$ variables and the data complexity will be $O\big(n^d \cdot \log n\big)$.
\end{example}

For threshold queries, however, low tree-width is not sufficient. The reason is that it is already hard to decide if the number of results of an acyclic CQ is above a threshold (represented in binary).
\begin{propositionrep}\label{prop:hardness-largecounter}
  Given an acyclic conjunctive query $q$, a threshold $k$ in binary representation, and a database $D$, testing if $q$ returns at least $k$ tuples on $D$ is \conp-hard.
\end{propositionrep}
\begin{proof}
We reduce from the well-known \textsf{VALIDITY} problem. Consider a Boolean formula $\varphi$ in disjunctive normal form with clauses $c_1, c_2, \dots, c_m$ over variables $x_1, x_2, \dots, x_n$. \textsf{VALIDITY} asks if $\varphi$ is \emph{valid}, i.e., if $\varphi$ evaluates to true under every possible truth assignment $\alpha : \{x_1,\ldots,x_n\} \to \{\true,\false\}$.

\begin{figure}
\tikzset{ex/.style={inner sep=2.5pt}}
\tikzset{fr/.style={draw,inner sep=2.5pt}}
\tikzset{tl/.style={draw,rounded corners,inner sep=2.5pt}}
\tikzset{output/.style={draw,rounded corners,double,inner sep=2.5pt}}
\begin{tikzpicture}[inner sep=4pt,node distance=15mm]

\node[] (t) {$t$}; 
\node[below of=t] (f) {$f$}; 

\node[right of=t,node distance=3cm,yshift=5mm] (c1) {$c_1$};
\node[below of=c1] (c2) {$c_2$}; 
\node[below of=c2] (c3) {$c_3$}; 

\draw (c1) edge[->] node[above,sloped] {$X_2,X_3,X_4$} (t);
\draw (c1) edge[->] node[above,sloped,pos=0.3] {$X_1,X_4$} (f);
\draw (c2) edge[->] node[above,sloped,pos=0.2] {$X_1,X_3$} (t);
\draw (c2) edge[->] node[above,sloped] {$X_1,X_2,X_4$} (f);
\draw (c3) edge[->] node[above,sloped,pos=0.2] {$X_1,X_4$} (t);
\draw (c3) edge[->] node[above,sloped] {$X_2,X_3,X_4$} (f);
\begin{scope}[xshift=7cm,yshift=-1cm,node distance=10mm]
  \node[tl] (u) {$u$};
  
  \node[output] (v1) at (-2,1.2)  {$v_1$};
  \node[output] (v2) at (-2,.4)  {$v_2$};
  \node[output] (v3) at (-2,-.4)  {$v_3$};
  \node[output] (v4) at (-2,-1.2)  {$v_4$};
  
  \draw 
  (u) edge[->] node[above,pos=0.7] {$X_1$} (v1)
  (u) edge[->] node[above,pos=0.7] {$X_2$} (v2)
  (u) edge[->] node[above,pos=0.7] {$X_3$}(v3)
  (u) edge[->] node[above,pos=0.7] {$X_4$} (v4)
  ;
\end{scope}
\end{tikzpicture}    

\medskip
$(\lnot x_1 \land  x_2 \land x_3) \lor (\lnot x_2 \land x_3 \land \lnot x_4) \lor (x_1 \land \lnot x_2 \land \lnot x_3)$
    \caption{Construction of the Property Graph (left) and query (right) for coNP-Hardness}
    \label{fig:sat-pg}
\end{figure}

We will define a database $D$, a conjunctive query $q$, and a threshold $k$ such that $q$ returns at least $k$ answers on $D$ iff $\varphi$ is valid.
The database $D$ uses the binary relations $X_1,\ldots,X_n$, has $\adom(D) = \{t, f\}\cup \{c_1, c_2, \dots, c_m\}$, and is illustrated in Figure~\ref{fig:sat-pg}. More formally, we put $(c_j,t)$ in relation $X_i$ iff assigning $x_i =  \true$ \emph{is compatible with} $c_j$, i.e., does not falsify clause $c_j$. Similarly, we put $(c_j,f)$ in $X_i$ iff assigning $x_i = \false$ is compatible with $c_j$, i.e.,  does not falsify $c_j$. 
Consider the CQ
\begin{equation*}
    q(v_1, v_2, \dots , v_n) = \exists u \; \bigwedge_{i=1}^n X_i(u,v_i)\,.
\end{equation*}
Notice that $\varphi$ is valid iff query $q$ at least $2^n$ tuples over $D$. Intuitively, 
each truth assignment $\alpha$ of $x_1,\ldots,x_n$ corresponds to a binding of $v_1,\ldots,v_n$ to the constants $t$ and $f$, where we map $v_i$ to $t$ iff $\alpha(x_i) = \true$. The condition $\wedge_{i=1}^n X_i(u,v_i)$ then tests if there exists a clause that is satisfied by $\alpha$: we can map $u$ to $c_i$ iff the clause $c_i$ is satisfied by $\alpha$.
\end{proof}
So, we cannot have a polynomial-time algorithm even for evaluating Boolean acyclic threshold queries of the form $\exists^{a,b} \bar y \, p(\bar y)$, unless $\ptime = \np$. This is why one focus in the paper is on
\begin{center}
    \emph{pseudopolynomial-time} algorithms\\ for threshold queries of low tree-width.
\end{center}
We call an algorithm \emph{pseudopolynomial}, if it is a polynomial-time algorithm assuming that the numerical values $a$ and $b$ in ``$\exists^{a,b} \bar y$'' are represented in unary (instead of binary). For instance, a pseudopolynomial algorithm can evaluate threshold queries of the form $ \exists^{a,b} \bar y\, p(\bar y)$ by keeping $b+1$ intermediate results in memory, which is not possible in a polynomial-time algorithm.

\newstuff{
Let us revisit Example~\ref{ex:boolean} and rewrite the query in a suitable way to produce its output. The next rewriting is inspired by research on constant-delay enumeration and answer counting for CQs.
\begin{example} \label{ex:enumeration}
The rewriting in Example~\ref{ex:boolean} is not suitable for non-Boolean evaluation because it projects out answer variables. 
The only way to rewrite $q$ while keeping track of all answer variables is 
\begin{align*}
    q''(x,y,z) \leftarrow &  \mathit{Assets}(x,y), U(x,z)\,;\\ \quad U(x,z) \leftarrow & \mathit{Subsidiary}(w,x), \mathit{Shareholder}(w,z)\,.
\end{align*}
The standard approach for constant-delay enumeration algorithms \cite{BaganDG07} first has a preprocessing phase, in which it
materializes $U$ and groups $\mathit{Assets}$ and $U$ by $x$. In the enumeration phase it iterates over possible values of $x$ and, for each value of $x$, over the contents of the corresponding groups of $\mathit{Assets}$ and $U$. The  complexity of the preprocessing phase is then affected by the cost of materializing $U$, which can be quadratic in the worst case. Overall, the complexity is $O(n^2 \cdot \log n)$ over databases of size $n$.
\end{example}
%
Evaluating a threshold query is closely related to constant-delay enumeration and counting answers to CQs. Indeed, a threshold query of the form $\exists^{a,b} \bar y \, p(\bar y)$ can be evaluated by enumerating answers to $p$ up to threshold $b+1$ or by counting all answers to $p$. For both these tasks, however, tractability relies on more restrictive parameters of the query. 
For enumeration, tree-width needs to be replaced with its \emph{free-connex} variant \cite{BaganDG07}, which treats answer variables in a special way.  
For counting, the additional parameter is the \emph{star-size} \cite{DurandM15}. Intuitively, it measures how many answer variables are maximally connected to a non-answer variable. The query $q$ has star-size $2$ because the existentially quantified variable $w$ is connected to two answer variables, $x$ and $z$.}

Thus, in the general approaches to constant-delay enumeration and counting, complexity is very sensitive to the interaction between answer and non-answer variables. 
Our key insight is that in the presence of a threshold this is no longer the case and low tree-width is sufficient.

\begin{example}\label{ex:threshold}
Consider again query $q$ from Example~\ref{ex:boolean} and suppose that we should return up to $c$ answers. 
We can rely on the rewriting in Example~\ref{ex:boolean}, but we need to store additional information when materializing the views. For each $w$ in $V$ we store up to $c$ witnessing values of $z$ such that $\mathit{Shareholder}(w,z)$ holds. 
Similarly, for each $x$ in $U$ we store up to $c$ values of $z$ that were stored as witnesses for some $w$ in $V$ with $\mathit{Subsidiary}(w,x)$. 
Now, we can obtain up to $c$ answers to $q$ by taking the join of $\mathit{Assets}(x,y)$ with $U(x)$ and iterating through the witnessing values of $z$ for each $x$.
Both extended materialization steps, as well as the final computation of answers, can be realized in time $O\big(c\cdot n \cdot \log(c\cdot n)\big)$. 
If we are to count answers up to threshold $c$, we can just count the ones returned by the algorithm above. 
\end{example}

This idea allows evaluating low tree-width TQs of the form $\exists^{a,b}\bar y\, p(\bar y)$ in pseudopolynomial-time. It is also crucial in our treatment of general TQs of the form $q(\bar x) \land \exists^{a,b}\bar y\, p(\bar x, \bar y)$ but, as the following proposition shows, we cannot expect pseudopolynomial evaluation algorithms even for \emph{acyclic} threshold queries. Our proof uses a reduction from \textsf{MINSAT}~\cite{KohliKM94}.

\begin{propositionrep}
\label{prop:hardness}\leavevmode
Boolean evaluation of acyclic at-least and at-most threshold queries is \np-hard, even if thresholds are given in unary.
\end{propositionrep}
\begin{proof} 
    For at-most queries, we reduce from \textsf{MAXSAT}. 
    The database $D$ is constructed in the same way as in Proposition~\ref{prop:hardness-largecounter}, except that we add a unary relation $\node(\cdot)$ that contains all elements in $\adom(D)$.
    Boolean evaluation of the at-most query
    \[t(v_1,\ldots,v_n) = \bigwedge_{i=1}^n \node(v_i) \land \exists^{\leq k} u\; \bigwedge_{i=1}^n X_i(u, v_i)\]
    checks if there exists an assignment that satisfies at most $k$ clauses in the DNF formula $\varphi$.
    For the dual CNF formula, this means that there is an assignment that falsifies at most $k$ of its clauses. This means that this assignment satisfies at least $m-k$ of its clauses. 

    For at-least queries, we reduce from \textsf{MINSAT}, which is also \np-complete \cite{KohliKM94}.
    Boolean evaluation of the at-least query
    \[t(v_1, \ldots v_n) = \bigwedge_{i=1}^n \node(v_i) \land \exists^{\geq k} u\; \bigwedge_{i=1}^n X_i(u, v_i)\]
    checks if there is an assignment that satisfies at least $k$ clauses in the DNF formula $\varphi$. 
    For the dual CNF / SAT formula, this means that there is an assignment that falsifies at least $k$ of its clauses. 
    This means that this assignment satisfies at most $m-k$ of its clauses. 
\end{proof}

The reason is that acyclic queries can have arbitrarily high  \emph{free-connex tree-width}, which is the actual source of hardness. 
For TQs of \emph{bounded free-connex tree-width} our approach will yield pseudopolynomial evaluation algorithms for all variants (E1)--(E5). That is, our results are tight in terms of combined complexity. 

In the remainder of the paper, we will analyze algorithms using $\tilde O$-notation. We will use this notation to reflect the \emph{data complexity} of the algorithms and to hide \emph{logarithmic factors}. Essentially, using $\tilde O$ allows us to freely use sorting and indexes such as B-trees. For instance, if we say that something can be done in time $\tilde O(n^2)$ we mean that its data complexity is in time $O(n^2 \log n)$.
\section{Threshold Queries in Theory}
\label{sec:technical-results}



In this section we define the notions of widths and decompositions informally discussed in Section~\ref{sec:exploiting-thresholds}, explore in depth the approach to threshold queries via exact counting, develop the idea illustrated in Example~\ref{ex:threshold} to cover arbitrary CQs in a slightly more general setting involving grouping, and employ the obtained algorithm to construct a single data structure that supports constant-delay enumeration, counting, and sampling answers to TQs.

\subsection{Tree decompositions and how to find them}
\label{ssec:decompositions}

\begin{figure}
\centering
\begin{tikzpicture}[>=stealth',semithick]
\footnotesize
\tikzset{slim/.style={inner sep=1pt}}
\begin{scope}[xshift=0cm,yshift=0cm]
  \node at (-0.8,0) {$T_1$};
  \node[slim] (A) at (0,0) {$\{x,y\}$};
  \node[slim] (B) at (0,-0.5) {$\{w,x\}$};
  \node[slim] (C) at (0,-1) {$\{w,z\}$};
  \draw (A) -- (B);
  \draw (B) -- (C);
\end{scope}

\begin{scope}[xshift=0cm,yshift=-1.5cm]
  \node at (-0.8,0) {$T_2$};
  \node[slim] (A) at (0,0) {$\{x,y,z\}$};
  \node[slim] (B) at (0,-0.5) {$\{w,x,z\}$};
  \draw (A) -- (B);
\end{scope}

\begin{scope}[xshift=2.5cm,yshift=0cm]
  \node[slim] at (-1,0) {$T_3$};   
  \node[slim] (y) at (0,0) {$\{x\}$};
  \node[slim] (xy) at (-0.5,-0.5) {$\{x,y\}$};
  \node[slim] (x) at (-1,-1) {$\{y\}$};      
  \node[slim] (xu) at (-0.5,-1.5) {$\{y, u\}$};
  \node[slim] (xus) at (-1,-2) {$\{y, y',u\}$};
  \node[slim] (yz) at (0.5,-0.5) {$\{x,z\}$};
  \node[slim] (z) at (1,-1) {$\{z\}$};   
  \node[slim] (zv) at (0.5,-1.5) {$\{z, v\}$};
  \node[slim] (zvt) at (1,-2) {$\{z,z',v\}$};

  \draw (y) edge[-] (xy);
  \draw (xy) edge[-] (x);
  \draw (x) edge[-] (xu);
  \draw (xu) edge[-] (xus);
  \draw (y) edge[-] (yz);
  \draw (yz) edge[-] (z);
  \draw (z) edge[-] (zv);
  \draw (zv) edge[-] (zvt);
\end{scope}

\begin{scope}[xshift=6cm,yshift=0cm]
  \node[slim][slim] at (-1,0) {$T_4$};
  \node[slim] (y) at (0,0) {$\{x\}$};
  \node[slim] (xy) at (-0.5,-0.5) {$\{x,y\}$};
  \node[slim] (x) at (-1,-1) {$\{y\}$};   
  \node[slim] (xu) at (-0.5,-1.5) {$\{y, y'\}$};
  \node[slim] (xus) at (-1,-2) {$\{y', u\}$};
  \draw (y) -- (xy);
  \draw (xy) -- (x);
  \draw (x) -- (xu);
  \draw (xu) -- (xus);
  \node[slim] (yz) at (0.5,-0.5) {$\{x,z\}$};
  \node[slim] (z) at (1,-1) {$\{z\}$};  
  \node[slim] (zv) at (0.5,-1.5) {$\{z, z'\}$};
  \node[slim] (zvt) at (1,-2) {$\{z',v\}$};
  \draw (y) -- (yz);
  \draw (yz) -- (z);
  \draw (z) -- (zv);
  \draw (zv) -- (zvt);
\end{scope}
\end{tikzpicture}
\caption{Tree decompositions}
\label{fig:decompositions}
\end{figure}

\OMIT{ 
\begin{figure}
\centering
\begin{subfigure}{.35\linewidth}
\centering
\begin{tikzpicture}[node distance=0.8cm]
    \node(A) {$\{x,y\}$};
    \node (B) [below of = A] {$\{w,x\}$};
    \node (C) [below of = B] {$\{w,z\}$};
    \draw (A) -- (B);
    \draw (B) -- (C);
\end{tikzpicture}
\caption[short]{A decompostion.}
\label{fig:ptree}
\end{subfigure}
\quad
\begin{subfigure}{.55\linewidth}
\centering
\begin{tikzpicture}[node distance=8mm]
    \node (A) {$\{x,y,z\}$};
    \node (B) [below of = A] {$\{w, x, z\}$};
    \draw (A) -- (B);
\end{tikzpicture}
\vspace{8mm}
\caption[short]{A free-connex  decomposition.}
\label{fig:qtree}
\end{subfigure}
\caption{Tree decompositions of the query $q(x,y,z) \leftarrow \mathit{Assets}(x,y), \mathit{Subsidiary}(w,x), \mathit{Shareholder}(w,z)$.}
\end{figure}
}

The rewritings discussed in Section~\ref{sec:exploiting-thresholds} are guided by tree decompositions of queries. 
A \emph{tree decomposition} of a conjunctive query $q$ is a finite tree $T$ with a set $X_v \subseteq \var(q)$, called a \emph{bag}, 
assigned to each node $v$ of $T$, satisfying the following conditions:
\begin{enumerate}
    \item 
    for each atom $A$ of $q$ there exists a node $v$ of $T$ such that $\var(A) \subseteq X_v$ (we say that $v$ \emph{covers} $A$);
    \item 
    for each variable $x \in \var(q)$, the set of nodes $v$ of $T$ such that $x \in X_v$ forms a connected subgraph of $T$.
\end{enumerate} 
By the \emph{width of $T$} we shall understand $\max_{v\in T}|X_v|$.\footnote{It is customary to define the width of decomposition $T$ as $\max_{v\in T}|X_v| - 1$, to ensure that tree-shaped queries have tree-width 1. For the purpose of this paper we prefer not to do it, thus avoiding adjustments by 1 in multiple formulas.} 
The \emph{tree-width of query $q$}, written as $\tw(q)$, is the minimal width of a tree decomposition of $q$. 
For example, $T_1$ in Figure~\ref{fig:decompositions} is a tree decomposition of width 2 for the query $q$ in Example~\ref{ex:boolean}. Since $q$ contains atoms involving two variables, it does not admit a tree decomposition of width 1. Hence, $\tw(q) = 2$.

A tree decomposition $T$ of a query $q$ is \emph{$X$-connex} for a set $X \subseteq \var(q)$, if there exists a connected subset $U$ of nodes of $T$, containing the root of $T$, such that the union of bags associated to nodes in $U$ is precisely $X$. Note that there is exactly one such $U$ that is maximal: it is the one that includes all nodes $u$ with $X_u \subseteq X$. We shall refer to it as \emph{the maximal $X$-connex set in $T$}.
The \emph{$X$-connex tree-width} of $q$ is the minimal width of an $X$-connex tree decomposition of $q$. If $X = \fvar(q)$, we speak of \emph{free-connex decompositions} and \emph{free-connex tree-width}; we write $\fctw(q)$ for the free-connex tree-width of $q$. For instance, $T_2$ in Figure~\ref{fig:decompositions} is a free-connex tree decomposition of width 3 for the query $q$ of Example~\ref{ex:boolean}. 
It is not hard to see that $q$ has no free-connex decomposition of width 2. That is,  $\tw(q) = 2$ but $\fctw(q) = 3$. This difference can be arbitrarily large, e.g., for the CQs we used in the proof of Proposition~\ref{prop:hardness-largecounter}.

A tree decomposition $T$ of $q$ is \emph{$X$-rooted}, for $X \subseteq \var (q)$, if the root bag of $T$ is exactly  $X$. The \emph{$X$-rooted tree-width of $q$} is the minimal tree-width of an $X$-rooted tree decomposition. By analogy to free-connex, if $X=\fvar(q)$, we speak of \emph{free-rooted decompositions} and \emph{free-rooted tree-width}.

It is convenient to work with tree decompositions $T$ of a special shape. 
A node $u$ of $T$ is: a \emph{project} node if it has exactly one child $v$ and $X_u \subsetneq X_v$; a \emph{join} node if it has exactly two children, $v_1$ and $v_2$, and $X_u = X_{v_1} \cup X_{v_2}$. 
We say that $u$ is \emph{safe} if each variable in $X_u$ occurs in an atom of $q$ that is covered either by $u$ or by a descendant of $u$. 
We say that $T$ is \emph{nice} if each node of $T$ is either a leaf or a project node or a join node, and all nodes of $T$ are safe.

\begin{lemmarep}\label{lem:nice}
Each tree decomposition $T$ of a conjunctive query $q$ can be transformed in polynomial time into a nice tree decomposition $T'$ of the same width and linear size. 
Moreover, if $T$ is $X$-rooted or $X$-connex, so is $T'$.
\end{lemmarep}

\begin{proof}
We perform the transformation in several steps. 
The first step is to ensure that $T$ is safe. This is done by pulling variables up the tree. Consider an unsafe node $u$, whose descendants are all safe. Let $Y_u$ be the set of variables in $X_u$ that violate the safety condition. Because all descendants of $u$ are safe, it follows immediately that they do not contain variables from $Y_u$. Consequently, after removing $Y_u$ from $X_u$, we still have a tree decomposition of $q$: the connectedness condition holds and each bag covers the same atoms it covered before the modification. This also means that no safe nodes became unsafe. The node $u$ itself, however, is now safe. Proceeding this way we make all nodes of $T$ safe. 

The second step is to ensure that each internal node $u$ either is a project node or contains in its bag the union of the bags of its children. This is done simply by inserting, between each two nodes, an intermediate project node, whose bag is the intersection of the bags of its parent and its unique child. This does not affect safety. 

The third step is to ensure that each internal node either is a project node or its bag is \emph{equal} to the union of the bags of its children. Consider an internal node $u$ such that $X_u$ is a strict superset of the union $Y_u$ of the bags of the children of $u$. Because $u$ is safe, for each variable $x \in X_u \setminus Y_u$, there exists an atom $A_x$ of $q$ that uses $x$ and is covered either by $u$ or by one of the descendants of $u$. It cannot be covered by a strict descendent of $u$, because these do not contain $x$. Hence, $A_x$ is covered by $u$. Let us add a new child $v$ of $u$ and let $X_v = \bigcup_{x \in X_u \setminus Y_u} \var(A_x)$. Note that $X_u \setminus Y_u \subseteq X_v \subseteq X_u$, so $X_u$ now is the union of the bags of its children. The new node $v$ is a leaf and it is clearly safe. It should also be clear that the addition of $v$ does not break $T$ in any way: it is still a tree decomposition of $q$, still satisfies all previously ensured properties. Proceeding this way we fix all nodes of $T$ that need fixing. 

The final step is to ensure that all internal nodes are either project nodes or join nodes. We already ensured that for each internal node $u$ that is not a project node, $X_u$ is the union of the bags of the children of $u$. The only issue now is that $u$ can have only one child or more than two children. If $u$ has only one child $v$ then $X_u = X_v$, so we can remove $v$ and promote all its children to become children of $u$, without affecting the guarantees we already achieved. If $u$ has more than two children, we can fix it by introducing between $u$ and its children a complete binary tree of join nodes, of logarithmic depth and linear size.

It is routine to verify that all four steps  preserve the property of being $X$-rooted and the property of being $X$-connex.
\end{proof}

\newstuff{We now introduce some notions that will be useful later.}  With each node $u$ in a tree decomposition $T$ of $q$ we associate the subquery $q_u$ obtained by taking all atoms of $q$ over the variables appearing in the subtree of $T$ rooted at $u$, and quantifying existentially all used variables except those in $X_u \cup \fvar(q)$. Let $\check q_u$ be the full CQ obtained by taking all atoms of $q_u$ that do not occur in $q_{v_1}\land q_{v_2} \land \dots \land q_{v_k}$, where $v_1, v_2, \dots, v_k$ are the children of $u$; we then have   $\var(\check q_u)= \fvar(\check q_u)\subseteq X_u$. 
For instance, for the query $q$ discussed in Section~\ref{sec:exploiting-thresholds} and its tree  decomposition $T_1$ shown in Figure~\ref{fig:decompositions} with nodes numbered 0, 1, 2 starting from the root, we have $q_0 = q = \exists w \: \mathit{Assets}(x,y) \land \mathit{Subsidiary}(w,x) \land \mathit{Shareholder}(w,z)$, $\check q_0 = \mathit{Assets}(x,y)$, $q_1= \exists x\, \mathit{Subsidiary}(w,x) \land \mathit{Shareholder}(w,z)$,  $\check q_1 = \mathit{Subsidiary}(w,x)$, and $q_2=\check q_2 = \mathit{Shareholder}(w,z)$.
In general, if $u$ is a leaf then $q_u=\check q_u$ and if $u$ is the root then $q_u = q$. 
One can evaluate $q$ on a database $D$ by computing $q_u(D)$ bottom up, as follows. Assuming that $T$ is nice, if $u$ is a leaf then 
\[q_u(D) = \check q_u (D)\,,\]
if $u$ is a project node with child $v$ then \[q_u(D) = \pi_{\fvar(q_u)}\big(q_{v}(D)\big)\,, \] and if $u$ is a join node with children $v_1$ and $v_2$ then \[q_u(D) = \check q_u(D) \bowtie  q_{v_1}(D) \bowtie q_{v_2}(D)\,.\] 
If $T$ is free-rooted then $\fvar(q_u) \subseteq X_u$ for each $u$ and the above computation can be performed in time $\tilde O(|D|^d)$, where $d$ is the width of $T$. 
The following simple fact will also be useful. 

\begin{lemmarep} \label{lem:safe}
Let $T$ be a nice tree decomposition of a CQ $q$. 
\begin{enumerate}
\item For each node $u$ in $T$ it holds that $X_u \subseteq \fvar(q_u)$.
\item If $T$ is $X$-connex and $U$ is the maximal $X$-connex set in $T$, then each node in $U$ either has all its children in $U$ or it is a $U$-leaf, that is, it has no children in $U$.
\end{enumerate}
\end{lemmarep}

\begin{proof}
The first item follows immediately from the safety of node $u$.
For the second item, take a node $u\in U$ that is not a $U$-leaf. Node $u$ cannot be a leaf in $T$, so it is either a join node or a project node. If $u$ is a join node, then the bags associated to its children are subsets of $X_u$, so the children of $u$ belong to $U$ by maximality. If $u$ is a project node, then it has exactly one child, and because $u$ is not a $U$-leaf, this child must belong to $U$. 
\end{proof}

Tree decompositions are not easy to find. Indeed, determining if an arbitrary graph admits a tree decomposition of width at most $d$ is NP-hard \cite{Arnborg87}. However, the problem has been studied in great depth and there is an ongoing effort of making these approaches practical at large scale. For instance, computing tree decompositions of large graphs was the topic of the PACE Challenge \cite{DellHJKKR16,DellKTW17} twice. 

However, in the present context, we only want to compute \emph{tree decompositions of queries}, which are very small in practice. There are libraries available \cite{FischlGLP19,detkdecomp} that can find optimal tree decompositions of queries very efficiently. Indeed, DetkDecomp~\cite{detkdecomp} was used to compute the tree-width of more than 800 million real-world queries \cite{bonifati2019navigating,BonifatiMT20,BonifatiMT-sigmod20} and worked very efficiently. Importantly for us, the analysis in \cite{bonifati2019navigating,BonifatiMT20,BonifatiMT-sigmod20} showed that real-life queries have very low tree-width.

Each algorithm for computing tree decompositions can be used also to compute $X$-rooted and $X$-connex tree decompositions, with quadratic overhead \cite{BaganDG07}. In our complexity estimations we rely on Bodlaender's algorithm \cite{Bodlaender96}, which allows computing optimal tree decompositions in linear time (assuming bounded tree-width).


\subsection{Threshold queries via exact counting}
\label{ssec:exact-counting}

Processing a threshold query  \[t(\bar x) = q(\bar x) \land \exists^{a,b} \bar y\, p(\bar x, \bar y)\] involves \newstuff{counting, for each answer $\eta$ in $q(D)$, how many answers in $p(D)$ are compatible to $\eta$. Formally, this means that we need to determine 
the size of $p(D, \eta)$ for each  $\eta \in q(D)$, where $p(D, \eta)$ is the set of answers in $p(D)$ that are compatible to $\eta$}. So we need to solve the following computational problem for $p$.
\begin{center}
\fbox{
\begin{minipage}{.8\linewidth}
\emph{Counting answers to $p$ grouped by $X \subseteq \fvar(p)$ over database $D$} consists in computing all pairs $(\eta, k)$ such that $\eta : X \to \adom(D)$ and $k = |p(D,\eta)|$.
\end{minipage}
}
\end{center}

Counting answers can leverage low \emph{star-size}~\cite{DurandM15}. While the original notion is designed to fit hypertree decompositions, we shall work with a slightly faster growing variant that fits tree decompositions better and is much easier to define; the two variants coincide for queries using at most binary atoms. 
The \emph{star-size} of a conjunctive query $q$, written $\ssz(q)$, is the least positive integer $f$ such that by grouping atoms and pushing quantifiers down, we can rewrite $q$ as $q_1 \land q_2 \land \dots \land q_\ell$ with $|\fvar(q_i)|\leq f$ or $\qvar(q_i) = \emptyset$ for all $i$.  The following is a routine generalization of the result on counting answers \cite{DurandM15}. 

\begin{propositionrep} 
\label{prop:counting-starsize}
Counting answers grouped by $X$ for conjunctive queries of $X$-rooted tree-width $d$ and star-size $f$ over databases of size $n$ can be done in time $\tilde O\big(n^{d\cdot f}\big)$.
\end{propositionrep}
\begin{proof} 
Let $q$ be a conjunctive query of tree-width $d$ and star-size $f$, and let $X \subseteq \fvar(q)$ be a set of grouping variables. Let $T$ be an $X$-rooted tree decomposition of $q$, of width $d$. 
Because $\ssz(q) = f$, we can equivalently write $q$ as the conjunction of CQs $q_1, q_2, \dots, q_\ell$, where $|\fvar(q_i)| \leq f$ or $\qvar(q_i)=\emptyset$ for each $i$. Without loss of generality we can assume that $\qvar(q_i)$ are pairwise disjoint 
and if $\qvar(q_i)=\emptyset$ then $q_i$ consists of a single atom. 
Let us rewrite $q$ as $q' = V_1(\bar x_1) \land V_2(\bar x_2) \land \dots \land V_\ell(\bar x_\ell)$ using views $V_i(\bar x_i) \gets q_i(\bar x_i)$ where, for each $i$,  the tuple $\bar x_i$ lists each variable from $\fvar(q_i)$ exactly once.
We can transform $T$ into an $X$-rooted tree decomposition $T'$ of $q'$, by replacing each occurrence of  $x\in \qvar(q_i)$ with the whole set $\fvar(q_i)$ for each $i$. Because the sets $\qvar(q_i)$ are pairwise disjoint, $T'$ has width $d\cdot f$. Using Lemma~\ref{lem:nice} we can assume that $T'$ is nice. 
 
For each $u$ in $T'$, let $X'_u$ be the associated bag and let $q'_u$ be the subquery of $q'$ obtained by dropping all atoms that use a variable not occurring in the subtree of $T'$ rooted at node $u$. We have $X'_u \subseteq \fvar(q'_u)=\var(q'_u)$ for all $u$.
We process $T'$ bottom-up and, for every node $u$, we compute the set $R_u$ of all records of the form $(\eta, k)$ where  $\eta : X'_{u} \to \adom(D)$ and $k = |q'_u(D, \eta)|$. That is, $R_u$ is the solution to the problem of counting answers to $q'_u$ grouped by $X'_u$ over $D$. We also interpret a record $(\eta,k)$ as the binding extending $\eta$ to a fresh variable $\mathit{witnessCount}$ with value $k$. Note that if $u$ is the root of $T'$, then $q'_u = q'$, and the whole algorithm can return $R_u$. It remains to show how to compute $R_u$, assuming that $R_v$ is already known for all children $v$ of $u$. 

If $u$ is a leaf of $T'$, then $R_u = \big\{(\eta, 1) \bigm| \eta \in q'_u(D)\big\} \cup \big\{(\eta, 0) \bigm| \eta :X'_u \to \adom(D),  \eta \notin q'_u(D)\big\}$. It suffices to show that $q'_u(D)$ can be computed within the required time bounds. The query $q'_u$ is the conjunction of all atoms of the form  $V_i(\bar x_i)$ such that each variable in $\bar x_i$ belongs to $X'_u$. That is, evaluating $q'_u$ involves materializing the corresponding views and computing the join. If $q_i$ is a single atom, $V_i$ is already materialized. In the remaining case, $q_i$ has arity at most $f$ and, because it is a subquery of $q$, it has tree-width at most $d$. Every tree decomposition can be turned into a free-rooted tree decomposition by setting aside an arbitrary free variable $x$, adding all the remaining free variables to each bag, and taking any bag originally containing $x$ as the root; if the root holds some quantified variables as well, we add a bag with just free variables as its parent. Hence, $q_i$ has free-rooted tree-width at most $\delta = d+ f-1 \leq d \cdot f$; the latter inequality holds for all positive integers. Consequently, the corresponding view can be materialized in time $\tilde O\big(|D|^{\delta}\big)$. Because $|X'_u|\leq d\cdot f$, we can compute the join in time $\tilde O\big(|D|^{d\cdot f}\big)$. 

If $u$ is a project node with child $v$, then we let $R_u = \gamma_{X'_u; \mathrm{sum}(\mathit{witnessCount})}\big(R_v\big)$; that is, we group records $(\eta, k)$ in $R_v$ by $\pi_{X'_u}(\eta)$ and for each group return $\pi_{X'_u}(\eta)$ and the sum of all $k$'s in this group. This is easy to do in time $\tilde O\big(|D|^{d\cdot f}\big)$.

If $u$ is a join node with children $v_1$ and $v_2$, then $R_u$ can be obtained as the set of records $\big(\eta_1 \bowtie \eta_2, k\big)$ for
$(\eta_1,k_1)\in R_{v_1}$,  $(\eta_2,k_2)\in R_{v_2}$ such that $\eta_1$  is compatible with $\eta_2$ and \[k = \begin{cases} 
k_1\cdot k_2 & \text{ if } \eta_1 \bowtie \eta_2 \text{ satisfies all atoms of } q'_u \textrm{ that not present in } q'_{v_1} \land q'_{v_2}\,,\\
0 & \textrm{ otherwise}\,. 
\end{cases}\]
To compute $R_u$ we materialize the relevant views, as explained above, and then apply a variant of the merge-join algorithm. All this can be done in $\tilde O\big(|D|^{d\cdot f}\big)$.
\end{proof}



When we consider (constant-delay) \emph{enumeration algorithms}, we see that the state-of-the-art approaches, e.g., \cite{BaganDG07,IdrisUV17}, use a different parameter of the query. Instead of bounded \emph{star-size}, these rely on bounded \emph{free-connex tree-width}. At first sight, this difference is not surprising, because in the absence of a threshold, counting and enumeration cannot be reduced to each other.
But a closer look reveals that both parameters play a similar role: limiting them allows to reduce the problem to the much simpler case of full CQs by rewriting the input query as a join of views of bounded arity. 
This is readily visible for the star-size method, where the views correspond to the queries $q_i$ in the definition of star-size of $q$, but it is also true for the free-connex method: there, the views correspond to subtrees of the decomposition rooted at the shallowest nodes holding an existentially quantified variable. Hence, the methods can be used interchangeably and we can replace star-size with free-connex tree-width in Proposition~\ref{prop:counting-starsize}. Below, the \emph{$X$-rooted free-connex tree-width} of a query is the minimal width of a tree decomposition of the query that is both $X$-rooted and free-connex.


\begin{propositionrep} \label{prop:counting-freeconnex}
Counting answers grouped by $X$ for conjunctive queries of $X$-rooted free-connex tree-width $d$ over databases of size $n$ can be done in time $\tilde O\big(n^{d}\big)$. 
\end{propositionrep}

\begin{proof}
Let $q$ be a conjunctive query of $X$-rooted free-connex tree-width $d$. Let $T$ be a nice $X$-rooted free-connex decomposition of $q$ of width $d$ 
and let $U$ be the maximal free-connex set in $T$. Let $u_1, u_2, \dots, u_k$ be the $U$-leaves of $T$. By Lemma~\ref{lem:safe}, each node in $U$ either is a $U$-leaf or has all children in $U$. Consequently, each node in $T$ belongs either to $U$ or to the subtree $T_i$ of $T$ rooted at $u_i$, for some $i$.

With each $u_i$ we associate the subquery $q_i$ in the usual way; that is, we take the subquery of $q$ obtained by dropping atoms using variables that are not present in the subtree of $T$ rooted at $u$. By the definition of $U$ it follows that $\fvar(q_i) \subseteq X_{u_i} \subseteq \fvar(q)$.
We let $\check q$ be the full CQ obtained by dropping from $q$ all atoms occurring in $q_{1}\land q_{2} \land \dots \land q_{k}$. We can write $\check q$ as the conjunction of single atoms $q_{k+1}, q_{k+2}, \dots, q_{\ell}$.
Then, $q$ is equivalent to $q_{1}\land q_{2} \land \dots \land q_{\ell}$ and for each $i$ either $\qvar(q_i) = \emptyset$ and $q_i$ is a single atom or $|\fvar(q_i)|\leq|X_{u_i})| \leq d$. 
Thus, we are in a situation entirely analogous to the one in Proposition~\ref{prop:counting-starsize}, except that the role of star-size $f$ is now played by the $X$-rooted free-connex tree-width $d$. 

Continuing like in Proposition~\ref{prop:counting-starsize} we would obtain an algorithm evaluating $q$ on database $D$ with running time $\tilde O\big(|D|^{d^2}\big)$. But it is easy to improve this complexity. Let $q'= V_1(\bar x_1)\land V_2(\bar x_2)\land \dots \land V_\ell(\bar x_\ell)$ be the full CQ using views $V_i(\bar x_i) \gets q_i(\bar x_i)$, like in Proposition~\ref{prop:counting-starsize}. Then, $T$ can be turned into an $X$-rooted tree decomposition $T'$ of $q'$ simply by restricting it to $U$: for $i\leq k$ the atom $V_i(\bar x_i)$ is covered by node $u_i$ and for $i > k$ any node covering the underlying atom of $q$ will do. 
The width of $T'$ is at most $d$, rather than $d\cdot f$. Moreover, as $\fvar(q_i) \subseteq X_{u_i}$ for $i\leq k$, the subtree of $T$ rooted at $u_i$ can be turned into a free-rooted tree decomposition of $q_i$ simply by dropping variables not used in $q_i$ and adding $X_{u_i} \cap \fvar(q_i)$ on top as the new root. Hence, the free-rooted tree-width of $q_i$ is at most $d$, rather than $d + f - 1$. This means that the algorithm described in Proposition~\ref{prop:counting-starsize} will run in time $\tilde O(|D|^d)$.
\end{proof}

In particular, all answers to $p$ can be counted in  $\tilde O\big(n^{\tw(p)\cdot\ssz(p)}\big)$ by Proposition~\ref{prop:counting-starsize} or in 
$\tilde O\big(n^{\fctw(p)}\big)$ by Proposition~\ref{prop:counting-freeconnex}. The following lemma shows that the latter bound is tighter, so we shall rely on Proposition~\ref{prop:counting-freeconnex}. 

\begin{lemmarep} \label{lem:interplay}
For each conjunctive query $p$,
\[ \ssz(p) \leq \fctw(p) \leq \tw(p)\cdot \max\big(1, \ssz(p)\big) \,. \]
The same holds for the $X$-rooted variant and the $X$-connex variant, for every $X \subseteq \fvar(p)$. Moreover, there exist CQs $p$ with arbitrarily large $\fctw(p)$ that satisfy $\fctw(p)\leq \sqrt{\tw(p)\cdot\ssz(p)} +1$.
\end{lemmarep}

\begin{proof}
For the first inequality consider a free-connex decomposition of $p$ of width $d$. Let $T_1, T_2, \dots, T_k$ be the maximal subtrees of $T$ that do not contain a node holding free variables only. Then, for each $i$, all free variables of $q$ occurring in $T_i$ must be present in the root of $T_i$. Consequently, $T_i$ contains at most $d$ free variables. It follows that the star-size of $q$ is at most $d$. The reasoning is of course not affected when additional conditions on $T$ are imposed, like being $X$-rooted or $X$-connex for some $X \subseteq \fvar(p)$.

Consider now a tree decomposition $T$ of $p$ of width $d$. As we do not  assume that $T$ is free-connex, free variables can be located in disjoint subtrees of $T$. 
Let us write $p$ as $p_1 \land p_2\land \dots \land p_\ell$ such that, for each $i$, $p_i$ is connected and either  $|\fvar(p_i)| \leq \mathsf{ss}(p)$ or $\qvar(p_i) = \emptyset$. Let $T_i$ be the tree decomposition of $p_i$ induced by $T$; that is, the decomposition obtained from $T$ by removing all variables that do not occur in $p_i$.
Clearly, $T_i$ has width at most $d$, just like $T$. 
We now turn it into a free-rooted decomposition of $p_i$. First, we add to each bag of $T_i$ the whole set $\fvar(p_i)$ except one arbitrarily chosen variable. After this operation, some bag contains the whole set $\fvar(p_i)$. We reorganize the tree so that this bag becomes the root, and add $\fvar(p_i)$ on top as the new root. The resulting tree decomposition $T'_i$ has width at most $d + \ssz(p) - 1$. 

Let $T'$ be obtained from $T$ by including into each bag containing any variable from $\qvar(p_i)$ all variables from $\fvar(p_i)$ and then dropping all existentially quantified variables. Because the sets $\qvar(p_i)$ are pairwise disjoint, the sizes of bags in $T'$ are bounded by $d\cdot \mathsf{ss}(p)$.  
We claim that for each variable $x \in \fvar(p)$ the set of nodes holding  $x$ forms a connected subset of $T'$. Let $p_x$ be the conjunction of all queries $p_i$ that use $x$. Because each $p_i$ is connected, so is $p_x$. It follows that the set of nodes in $T$ that hold a variable from $p_x$ is connected, too. This is precisely the set of nodes of $T'$ that hold $x$. 

We now combine $T'$ with $T'_1, T'_2, \dots, T'_\ell$. For each $i$ there exists node $u_i$ in $T'$ that holds all variables in $\fvar(p_i)$: any node originally holding a variable from $\fvar(p_i)$ will do. The combined decomposition $T''$ is obtained by letting the root of $T'_i$ become a child of $u_i$.
It is routine to check that $T''$ is a free-connex decomposition of $p$ of width $d\cdot \ssz(p)$.
It is also straightforward to check that for each $X \subseteq \fvar(p)$, if $T$ is $X$-rooted or $X$-connex, so is $T''$.


Finally, for  \[q(x_1, \dots, x_n) = \exists x_0 \bigwedge_{i=0}^n \bigwedge_{j=0}^n  E(x_i, x_j)\]
we have $\ssz(q) = n$ and $\fctw(q) = \tw(q) = n+1$, so indeed $\fctw(q) \leq \sqrt{\tw(q) \cdot \ssz(q)} + 1$.
%
%
\end{proof}

Let us now come back to the threshold query $t$. By a tree decomposition of $t$ we shall mean a tree decomposition of the associated CQ $q \land p$.
Based on this, we define all variants of tree-width for threshold queries just like for CQs. Importantly, free-connex refers to $\fvar(t)$-connex, not $\fvar(q\land p)$-connex.
Applying Proposition~\ref{prop:counting-freeconnex} requires an $\fvar(p)$-connex decomposition; that is, $\ftvar(t)$-connex in terms of $t$, where $\ftvar(t) = \fvar(t) \cup \tvar(t)$.
Moreover, to avoid manipulating full answers, we need to factorize the evaluation of $q$ and counting answers to $p$ grouped by $\fvar(t)$ in a compatible way.
This can be done if our decomposition is also $\fvar(t)$-connex. Putting it together, we arrive at free-connex $\ftvar(t)$-connex decompositions of $t$.

\begin{figure}
    \centering
\begin{tikzpicture}[>=stealth']
\footnotesize
\tikzset{slim/.style={inner sep=1pt}}

\node[slim] (a1) at (0,0) {$a_1$}; 

\node[slim] (b1) at (-2.50,-0.5) {$b_1$}; 
\node[slim] (b2) at (-1,-0.5) {$b_2$}; 
\node[slim] (c1) at (0.00,-0.7) {$c_1$}; 
\node[slim] (c2) at (1,-0.5) {$c_2$}; 
\node[slim] (b3) at (2.50,-0.5) {$b_3$}; 

\node[slim] (d1) at (-1.75,-1) {$d_1$}; 
\node[slim] (d2) at (1.75,-1) {$d_2$}; 

\node[slim] (e1) at (-3.5,-.5) {$e_1$}; 
\node[slim] (e2) at (3.5,-.5)  {$e_2$}; 
\node[slim] (e3) at (4.5,-.5)  {$e_3$};

\draw[bend angle=25] (a1) edge[->,bend right] node[above, pos=0.7] {$B$} (b1);
\draw[bend angle=15] (a1) edge[->,bend right] node[above, pos=0.7] {$B$} (b2);
\draw (a1) edge[->] node[right,pos=0.35] {$C$} (c1);
\draw[bend angle=15] (a1) edge[->,bend left] node[above, pos=0.7] {$C$} (c2);
\draw[bend angle=25] (a1) edge[->,bend left] node[above, pos=0.7] {$B$} (b3);

\draw[bend angle=15] (b1) edge[->,bend right] node[above=1pt, pos=0.77] {$D$} (d1);
\draw[bend angle=15] (b2) edge[->,bend left] node[above, pos=0.66] {$D$} (d1);
\draw[bend angle=15] (c1) edge[->,bend left] node[above, pos=0.4] {$D$} (d1);

\draw[bend angle=15] (c1) edge[->,bend right] node[above, pos=0.4] {$D$} (d2);
\draw[bend angle=15] (c2) edge[->,bend right] node[above=1pt, pos=0.77] {$D$} (d2);
\draw[bend angle=15] (b3) edge[->,bend left] node[above, pos=0.65] {$D$} (d2);

\draw[bend angle=15] (d1) edge[->,bend left] node[below, pos=0.7] {$E$} (e1);
\draw[bend angle=14] (d2) edge[->,bend right] node[below, pos=0.8] {$E$} (e2);
\draw[bend angle=20] (d2) edge[->,bend right] node[below, pos=0.85] {$E$} (e3);
\end{tikzpicture}
\vspace{-3mm}
        \caption{Database from Example~\ref{ex:main}.}
    \label{fig:database}
\end{figure}

\begin{example}
\label{ex:main} 
Consider an at-least query 
\[t(x,y,z) = B(x,y) \land C(x,z) \land \exists^{\geq 3} (u,v) \; r(y, u) \land s(z,v) \]
for $r(y,u) = \exists y' \, D(y, y') \land E(y', u)$, $s(z,v) = \exists z' \, D(z, z') \land E(z', v)$. Figure~\ref{fig:decompositions} 
shows a free-connex $\ftvar(t)$-connex tree decomposition $T_3$ of $t$ of minimal width, which is 3.  
The subqueries $r(y,u)$ and $s(z,v)$ correspond to bags $\{y\}$ and $\{z\}$, respectively, and the subtrees rooted at these bags are $\{y\}$-rooted (resp. $\{z\}$-rooted) free-connex tree decompositions for these subqueries. 
Consider the input database in Figure~\ref{fig:database}.
Let us count the answers to $r(y,u)$ grouped by $y$ and the answers to $s(z,v)$ grouped by $z$ (using Proposition~\ref{prop:counting-freeconnex}) and store the resulting pairs in sets $R_{\{y\}}$ and $R_{\{z\}}$, respectively. 
We get $R_{\{y\}} = R_{\{z\}} = \big\{(b_1, 1), (b_2, 1), (b_3, 2), (c_1, 3), (c_2, 2)\big\}$, where a pair $(b,k)$ in $R_{\{y\}}$ means that for $y=b$ there are $k$ witnessing values of $u$, and similarly for $R_{\{z\}}$. 
This is the initial information that we shall now propagate up the tree. In the set $R_{\{x,y\}}$ we put triples $(a, b, k)$ such that $B(a,b)$ and for $y=b$ there are $k$ witnessing values of $u$; analogously for $R_{\{x,z\}}$. 
We get $R_{\{x,y\}} = \big\{(a_1,b_1, 1), (a_1,b_2, 1), (a_1,b_3, 2)\big\}$, $R_{\{x,z\}} = \big\{(a_1,c_1, 3), (a_1,c_2, 2)\big\}$.
These sets can be computed based on $R_{\{y\}}$ and $R_{\{z\}}$, respectively. The set $R_{\{x\}}$ stores pairs $(a,k)$ such that $k$ is the maximal number of witnessing pairs of values for $u$ and $v$ that any combination of $y=b$ and $z=c$ with $B(a,b)$ and $C(a,c)$ can provide. 
In our case, $R_{\{x\}} = \big\{(a_1,6)\big\}$.
Because $b$ and $c$ can be chosen independently from each other, we have $k= m \cdot n$, where 
$ m =\max_a\big\{ \ell \bigm | (a, b, \ell) \in R_{\{x,y\}}\big\}$ and 
$n=\max_b\big\{ \ell \bigm | (a, c, \ell) \in R_{\{x,z\}}\big\}$.
That is, the only information that needs to be passed from $\{x,y\}$ to $\{x\}$ is the set $R'_{\{x,y\}}$ of pairs $(a,m)$ with $m$ defined as above, and similarly for $\{x,z\}$. 
In our case, $R'_{\{x,y\}} = \big\{(a,2)\big\}$ and  $R'_{\{x,z\}} = \big\{(a,3)\big\}$. We return YES iff $R_{\{x\}}$ contains a pair $(a,k)$ with $k \geq 3$. In our case, it is so.
\end{example}

\begin{theoremrep}
\label{thm:poly}
Boolean evaluation of an at-most or at-least query $t$ of free-connex $\ftvar(t)$-connex tree-width $d$ over databases of size $n$ can be done in time $\tilde O(n^{d})$. The combined complexity of the algorithm is polynomial, assuming $d$ is constant. 
\end{theoremrep}

\begin{proof} 
The following proof is a simpler variant of the proof of Theorem~\ref{thm:pseudopoly}. The key difference in the data structure is that records contain only the witness count (no multiplicity) and for each binding we only store the maximal witness count for at-least queries and the minimal witness count for at-most queries. The answer then is YES iff the root contains a record with the witness count satisfying the bounds of the query. Below we spell out full details for the benefit of the reader wishing to see the proof of the present result without needing to go through the more complicated construction necessary for Theorem~\ref{thm:pseudopoly}.

Consider a database $D$ and a threshold query 
$ t(\bar x) = q(\bar x) \land \exists^{a,b} \, \bar y\: p(\bar x, \bar y)$. We assume that $t$ is either an at-least query or an at-most query; that is, either $b = \infty$ or $a = 0$.
Let $T$ be a nice free-connex $\ftvar(t)$-connex tree decomposition of $t$, of width $d$, and let $U$ be the maximal free-connex set in $T$.
Being a tree decomposition of $r(\bar x, \bar y) = q(\bar x) \land p(\bar x, \bar y)$, $T$ is also a tree decomposition of both $p$ and $q$ (up to dropping unused variables). Let $q_u$ and $p_u$ be the corresponding subqueries associated to node $u$. It follows that $r_u = q_u\land p_u$.
By Lemma~\ref{lem:safe}, we have $X_u \subseteq \fvar(r_u) = \fvar(q_u)\cup \fvar(p_u)$. If $u \in U$ then additionally $X_u \subseteq \fvar(t)$ and $\fvar (q_u) \subseteq \fvar (q) = \fvar(t)$, and consequently
$ X_u \subseteq \fvar(q_u) \cup \big (\fvar(p_u) \cap \fvar(t))$.
With each node $u \in U$ we associate the set $R_u$ of records that will provide information necessary for Boolean evaluation of $t$. A \emph{record for $u\in U$} is a pair $(\eta, k)$ consisting of a binding $\eta : X_u \to \adom(D)$ and a \emph{witness count} $k$ whose definition depends on whether $t$ is an at-least or an at-most query. If $t$ is an at-least query, then $k$ is the maximum of $|p_u(D, \eta')|$ with  $\eta'$ ranging over extensions of $\eta$ to $\fvar(q_u)\cup \big(\fvar(p_u) \cap \fvar(t)\big)$ such that  $\pi_{\fvar(q_u)}(\eta') \in q_u(D)$. If $t$ is an at-most query, then we replace maximum with minimum. In either case, if for some $\eta$ there is no $\eta'$, then there is no record for $\eta$ either.
We can interpret a record $(\eta, n, k)$ as a binding extending $\eta$ to a fresh variable $\mathit{witnessCount}$ with value $k$. 

By Lemma~\ref{lem:safe}, each node in $u$ either is a $U$-leaf or has all its children in $U$. Consequently, we can compute $R_u$ for $u \in U$ bottom-up. Let $u$ be a $U$-leaf. Then, by the definition of $U$, we have $X_u = \fvar(q_u) \cup  \big(\fvar(p_u) \cap \fvar(t)\big)$. Let $S_u$ be the solution to the problem of counting answers to $p_u$ grouped by $\fvar(p_u)\cap\fvar(t)$ over $D$. By restricting all bags to $\var(p_u)$ and adding $\fvar(p_u) \cap \fvar(t)$ on top as the new root, we turn the subtree of $T$ rooted at $u$ into a $\fvar(p_u) \cap \fvar(t)$-rooted  decomposition $T'$ of $p_u$ of width at most $d$. 

We claim that $T'$ is also $\fvar(p_u)$-connex. Indeed, recall that $T$ is $\ftvar(t)$-connex and let $V$ be the maximal $\ftvar(t)$-connex set in $T$. Let $V'$ comprise the  restriction of $V$ to the subtree rooted at $u$, as well as the newly added root of $T'$. Clearly, $V'$ is connected. Because $\fvar(p_u) = \ftvar(t) \cap \var(p_u)$, it follows that the union of bags stored in nodes from $V'$ is exactly $\fvar(p_u)$. This proves  the claim. 

Hence, we can use Proposition~\ref{prop:counting-freeconnex} to compute $S_u$ in time $\tilde O\big(|D|^d\big)$. The subtree of $T$ rooted at $u$ similarly induces an $\fvar(q_u)$-rooted decomposition of $q_u$ of width at most $d$, so we can compute the set $q_u(D)$ of all answers to $q_u$ over $D$ in time $\tilde O(|D|^d)$ as explained in Section~\ref{ssec:decompositions}. 
Then, we get $R_u$ as the join $R_u = q_u(D) \bowtie S_u$ defined as the set of records $(\eta_1 \bowtie \eta_2, k)$ such that $\eta_1 \in q_u(D)$,   $(\eta_2, k) \in S_u$, and $\eta_1$ and $\eta_2$ are compatible. 
This step can also be performed in $\tilde O\big(|D|^d\big)$. 

Suppose now that $u \in U$ is not a $U$-leaf. Then, all children of $u$ belong to $U$. Let us assume that their sets of records are already computed. 
If $u$ is a project node with child $v$, then $R_u$ is obtained from $R_v$ by grouping records $(\eta, k)$ in $R_v$ by $\pi_{X_u}(\eta)$ and returning for each group the binding  $\pi_{X_u}(\eta)$ and either the maximal witness count $k$ in the group if $t$ is a an at-least query, or the minimal witness count $k$ in the group if $t$ is an at-most query.
Because $|R_v| \leq |D|^d$, this step can also be done in time $\tilde O\big(|D|^d\big)$.

Suppose that $u$ is a join node with children $v_1$ and $v_2$.  First, we compute the set $R_{v_1,v_2}$ of records $\big(\eta_1 \bowtie \eta_2,  k_1\cdot k_2\big)$ 
such that $(\eta_1, k_1) \in R_{v_1}$, $(\eta_2,k_2) \in R_{v_2}$, and $\eta_1$ is compatible with $\eta_2$.
Next, we compute the \emph{semi-join} $R_{v_1,v_2, u} = R_{v_1,v_2} \ltimes \check q_u(D)$ 
of $R_{v_1,v_2}$ with $\check q_u(D)$; that is, the set of records $(\eta, k) \in  R_{v_1,v_2}$ such that $\pi_{\fvar(\check q_u)}(\eta) \in \check q_u(D)$. 
Finally, we compute
\[R_{u} = R_{v_1,v_2,u} \ltimes \check p_u(D) \cup R_{v_1,v_2,u} \triangleright \check p_u(D)\] 
as the union of $R_{v_1,v_2,u} \ltimes \check p_u(D)$ and the \emph{anti-join} $R_{v_1,v_2,u} \triangleright \check p_u(D)$ of $R_{v_1,v_2,u}$ with $\check p_u(D)$, defined as the set of records $(\eta, 0)$ for each  $(\eta,k) \in  R_{v_1,v_2,u}$ such that $\pi_{\fvar(\check p_u)}(\eta) \notin \check p_u(D)$.
Each of these operations can be performed in time $\tilde O\big(|D|^d\big)$.

Having computed $R_v$ for the root $v$ of $T$, we return YES iff $R_v$ contains a record $(\eta,k)$ with $a \leq k \leq b$.
\end{proof}
In terms of combined complexity, Theorem~\ref{thm:poly} is optimal as both conditions imposed on tree decompositions are needed. Indeed, the hard TQs used in Proposition~\ref{prop:hardness} have $\ftvar(t)$-connex tree-width 2; and the hard Boolean TQs stemming from Proposition~\ref{prop:hardness-largecounter} have free-connex tree-width 2.

\subsection{Threshold problems for CQs}
\label{ssec:threshold_cqs}
%
In terms of data complexity, however, we can improve Theorem~\ref{thm:poly}. Here we exploit the presence of thresholds to handle queries with unbounded $\ftvar(t)$-connex tree-width. We will cover not only at-least and at-most queries, but arbitrary TQs, and we will solve not only Boolean evaluation, but also constant-delay enumeration, counting answers, and sampling, all based on a single data structure. The small price we have to pay is pseudopolynomial combined complexity. 
Our starting point is again the problem of counting grouped answers, but this time up to a threshold. 

\begin{center}
\fbox{
\begin{minipage}{.95\linewidth}
\emph{Counting answers to $p$ grouped by $X \subseteq \fvar(p)$ up to a threshold $c$ over database $D$} consists in computing the set of pairs $(\eta, k)$ such that $\eta : X \to \adom(D)$ and $k = \min(c, |p(D,\eta)|)$.
\end{minipage}}
\end{center}
In the presence of a threshold, it is not impractical to solve counting by enumerating answers. This is what we shall do. 
\begin{center}
\fbox{
\begin{minipage}{.95\linewidth}
\emph{Computing answers to query $p$ grouped by $X \subseteq \fvar(p)$ up to a threshold $c$ over database $D$} consists in computing a subset $A$ of $p(D)$ that is \emph{complete for $X$ and $c$}; i.e., for each $\eta:X \to \adom(D)$ either 
$p(D,\eta) \subseteq A$ and $|p(D,\eta)|\leq c$, or $|p(D,\eta) \cap  A| = c$.
\end{minipage}}
\end{center}

Consider a CQ $p$,  a set $X\subseteq \fvar(p)$ of grouping variables, and a database $D$. We will use the term \emph{group} to refer to the set of answers to $p$ that agree on the grouping variables $X$; that is, a subset of $p(D)$ of the form $p(D, \eta)$ for some $\eta: X \to \adom(D)$. If the $X$-rooted tree-width of $p$ is $d$, then $|X| \leq d$ and the number of groups is $O(|D|^{d})$. Consequently, if we can compute answers to $p$ grouped by $X$ up to threshold $c$ in time $\tilde O(c \cdot |D|^{d})$, then we can also count them within the same time, because we will get at most $c$ answers per group. Additionally, for each $\eta:X \to \adom(D)$ with $p(D,\eta) = \emptyset$,
we need to include $(\eta, 0)$ into the result; this does not affect the complexity bound. 
Hence, it suffices to show how to compute answers grouped by $X$ up to threshold $c$. Having seen an illustrating example in Section~\ref{sec:exploiting-thresholds}, we are ready for the full solution. 

\begin{algorithm}[t]
	\caption{Answers to $p$ grouped by $X$ up to threshold $c$} 
	\label{algo:grouped-answers-threshold}
\begin{algorithmic}[1]
\State $T \gets $ a nice $X$-rooted tree decomposition of $p$
\Loop { through nodes $u$ of $T$ bottom-up}
    \If {$u$ is a leaf}   
	    $A_u \gets p_u (D)$ 
	\EndIf
	\If {$u$ is a project node with child $v$}  
	    \State $ A_u \gets \mathit{prune}_{X_u}^{\,c} \big(\pi_{\fvar(p_u)}(A_{v})\big)$
	\EndIf					
	\If {$u$ is a join node with children $v_1,v_2$} 
	    \State $A_u \gets \mathit{prune}_{X_u}^{\,c} \big (A_{v_1} \bowtie A_{v_2} \bowtie  \check{p}_u (D) \big )$	    
	\EndIf
\EndLoop
\end{algorithmic} 
\OMIT{	
\begin{algorithmic}[1]
\Function{$\mathit{prune}_{Y}^{c}$}{$A$} 
\State $B \gets \emptyset$;  $\xi  \leftarrow \mathit{null}$; $i \gets 0$;
\For {$\eta\in\mathbf{sort}_{Y}(A)$}
    \If{$\xi=\mathit{null}$ or $\xi \neq \pi_{Y}(\eta)$} 
        \State $\xi \gets \pi_{Y}(\eta)$; $i \gets 0$
    \EndIf
    \If{$i < c$}
        \State $B \gets B \cup \{\eta\}$; $i \gets i + 1$
    \EndIf
\EndFor
\State \textbf{return} $B$
\EndFunction
\end{algorithmic}   
}
\end{algorithm}

\begin{theoremrep} \label{thm:answers_upto}
Computing answers grouped by $X$ up to a threshold $c$ for conjunctive queries of $X$-rooted tree-width $d$ over databases of size $n$ can be done in time $\tilde O(c \cdot n^{d})$. 
\end{theoremrep}
\noindent
\begin{proofsketch}
Consider Algorithm~\ref{algo:grouped-answers-threshold}. We begin by computing a nice $X$-rooted tree decomposition $T$ of minimal width $d$, as described in Section~\ref{ssec:decompositions}.
That is, each bag of $T$ has size at most $d$ and the root bag is $X$. 
Consider the queries $p_u$ associated to nodes $u$ of $T$. By Lemma~\ref{lem:safe}, $X_u \subseteq \fvar(p_u)$.
By analogy to the evaluation algorithm described in Section~\ref{ssec:decompositions}, for each $u$ we solve the problem of computing answers to $p_u$ grouped by $X_u$ up to threshold $c$ over $D$; that is, we compute a subset $A_u \subseteq p_u(D)$ that is complete for $X_u$ and $c$. The final answer is then the set obtained in the root.  

If $u$ is a leaf, then $X_u = \fvar(p_u) = \var(p_u)$ and $p_u(D)$ itself is the only subset of $p_u(D)$ complete for $X_u$ and $c$. Because $|\var(p_u)|\leq d$, one can compute $p_u(D)$ in time $\tilde O\big(|D|^{d}\big)$.

Consider a project node $u$ and its unique child $v$. Then $X_u \subsetneq X_v$. To compute $A_u$ the algorithm projects $A_v$ on $\fvar(p_u)$ and then, using the operation $\mathit{prune}_{X_u}^{\,c}$, it groups the resulting set of bindings by $X_u$ and keeps only $c$ bindings from each group. It is clear that this can be done in time $\tilde O(c \cdot |D|^{d})$. 

Finally, consider a join node $u$ with children $v_1, v_2$. 
As explained in Section~\ref{ssec:decompositions}, the set $p_u(D)$ is then the join of $p_{v_1}(D)$, $p_{v_2}(D)$, and $\check p_u (D)$. We compute $A_u$ in exactly the same way, except that for each binding $\eta$ of variables in $X_u$ we only keep the first $c$ bindings extending $\eta$ and discard the remaining ones. Naive implementation takes time $\tilde O\big(c^2 \cdot|D|^d\big)$, but it is easy to modify the standard merge-join algorithm to achieve this in time $\tilde O\big(c \cdot |D|^{d}\big)$. 

It is routine to check that each $A_u$ is complete for $X_u$ and $c$.
\end{proofsketch}
\newstuff{Recall that $d$ is very small in practice and most queries can be expected to be acyclic~\cite{BonifatiMT20}. If we are just interested in returning answers up to a threshold $c$ (no grouping), then the algorithm underlying the present theorem improves the state-of-the-art algorithm from $\tilde{O}(n^f)$ to $\tilde{O}(c\cdot n)$ for acyclic queries, where $f$ is the \emph{free-connex treewidth} of the query, which can be large even for acyclic queries.}


\begin{proof}

It remains to prove that for each node $u$, the set $A_u$ computed by Algorithm~\ref{algo:grouped-answers-threshold} is complete for $X_u$ and $c$. We proceed by bottom-up induction on the nodes of the decomposition. The base case is immediate: if $u$ is a leaf, $A_u = p_u(D)$, so it is complete. 

Suppose that $u$ is a project node. Consider $\eta : X_u \to \adom(D)$. We need to check that $A_u$ contains sufficiently many bindings compatible with $\eta$. If for every binding $\eta'$ extending $\eta$ to $X_v$, we have that $p_v(D, \eta') \subseteq A_v$, then $p_u(D, \eta) \subseteq A_u$. Suppose that for some $\eta'$ extending $\eta$ to $X_v$ we have $p_v(D, \eta') \not\subseteq A_v$. Because $A_v$ is complete for $X_v$ and $c$ by the induction hypothesis, it must hold that $|p_v(D, \eta') \cap A_v| = c$. Because all bindings in $p_v(D, \eta') \cap A_v$ extend $\eta$ and their projections on $\fvar(p_u)$ are pairwise different (only variables from $X_v$ are projected out, and their values are fixed by $\eta'$ anyway), it follows that $|p_u(D, \eta) \cap A_u| = c$. Hence, $A_u$ is complete for $X_u$ and $c$.

Suppose now that $u$ is a join node. By the induction hypothesis, for each child $v$ of $u$,  $A_v$ is complete for $X_v$ and $c$. The interesting case is when $p_{v_i}(D, \eta) \not\subseteq A_{v_i}$ for some $v_i$ and some $\eta : X_{v_i} \to \adom(D)$. Will we not miss the discarded bindings now? As $A_{v_i}$ is complete for $X_{v_i}$ and $c$, we have $|p_{v_i}(D, \eta) \cap A_{v_i}|=c$. Consider a binding $\eta'$ extending $\eta$ to $X_u$. There are only two possibilities. If each of the remaining two joined sets contains at least one binding extending $\eta'$, the result of the join will contain at least $c$ such bindings. If some of the remaining joined sets contains no bindings extending $\eta'$, the result of the join will contain no such bindings either, and neither it would if we replaced each $A_{v_j}$ with $p_{v_j}(D)$ for each $j$. Thus, $A_u$ is complete for $X_u$ and $c$.

\end{proof}

\subsection{Processing threshold queries}

We now turn to processing general threshold queries. We give a unified approach to constant-delay enumeration, counting, and sampling, based on a single data structure that can be computed using the methods presented in Section~\ref{ssec:threshold_cqs}.

\begin{example} 
\label{ex:pseudopoly}
Consider again the database from Figure~\ref{fig:database} and the query $t$ from Example~\ref{ex:main}. This time we shall work with the free-connex tree decomposition $T_4$ of $t$ shown in Figure~\ref{fig:decompositions}; it has smaller width than $T_3$. Like before, the subqueries $r(y,u)$ and $s(z,v)$ correspond to bags $\{y\}$ and $\{z\}$, respectively, and the subtrees rooted at these bags are $\{y\}$-rooted (resp. $\{z\}$-rooted) tree decompositions for these subqueries, but they are not free-connex. We count the answers to $r$ grouped by $y$ and the answers to $s$ grouped by $z$, up to threshold 3, as described in Section~\ref{ssec:threshold_cqs}, and store the results in sets $R_{\{y\}}$ and $R_{\{z\}}$, respectively. Because the counts obtained in Example~\ref{ex:main} were all below $3$, $R_{\{y\}}$ and $R_{\{z\}}$ are just like before. Sets $R_{\{x,y\}}$ and $R_{\{x,z\}}$ are also computed like before. For Boolean evaluation we would now put into $R'_{\{x,y\}}$ all pairs $(a, m)$ such that $m$ is the maximal number of witnessing values $u$ that any $y=b$ with $B(a,b)$ can provide. To support constant-delay enumeration we need to pass more information up the tree:  we include all pairs $(a, k)$ such that some $y=b$ with $B(a,b)$ can provide $k$ witnesses (up to threshold 3); that is, we forget the values $b$, but we keep all values $k$ (up to 3), not only the greatest of them. 
In our case, $R'_{\{x,y\}} = \big\{(a_1,1), (a_1,2)\big\}$ and $R'_{\{x,y\}} = \big\{(a_1,2), (a_1,3)\big\}$.
In the root we store the set $R_{\{x\}}$ of values $a$ such that some combination of $y=b$ and $z=c$ with $B(a,b)$ and $C(a,c)$ can provide at least $3$ witnessing pairs of values for $u$ and $v$. The set $R_{\{x\}}$ can be obtained by taking all $a$ such that there exist  $(a,m) \in R'_{\{x,y\}}$ and $(a,n) \in R'_{\{x,z\}}$ with $m\cdot n \geq 3$. In our case, $R_{\{x\}} = \{a_1\}$. 

In the enumeration phase, we iterate over all combinations of $(a,m) \in R'_{\{x,y\}}$ and $(a,n) \in R'_{\{x,z\}}$ with $a \in R_{\{x\}}$. For each such combination we iterate over corresponding $(a,b,m) \in R_{\{x,y\}}$ and $(a,c,n)\in R_{\{x,z\}}$ 
and return $(a,b,c)$. 
In  our case this gives $(a_1,b_1,c_1)$, $(a_1,b_2,c_1)$, $(a_1,b_3,c_1)$, and $(a_1,b_3,c_2)$.
To access relevant $(a,b,m)$ and $(a,c,n)$ directly, we use an index that can be built while constructing $R'_{\{x,y\}}$ and $R'_{\{x,z\}}$ from $R_{\{x,y\}}$ and $R_{\{x,z\}}$

To support counting and sampling, we add multiplicities to the pairs stored in the nodes of the decomposition. Specifically, for each $(a, m) \in R'_{\{x,y\}}$ we also store the number of values $b$ with $B(a,b)$ that provide $n$ witnesses (up to 3). 
In our case, the multiplicity of $(a_1,1)$ in $R'_{\{x,y\}}$ is 2, and all other multiplicities in $R'_{\{x,y\}}$ and $R'_{\{x,z\}}$ are 1. 
Similarly, for each $a \in R_{\{x\}}$ we store the number of combinations of $b$ and $c$ with $B(a,b)$ and $C(a,c)$ that provide at least 3 pairs of witnesses. 
In our case, the multiplicity of $a_1$ in $R_{\{x\}}$ is $4$: the witnessing combinations are $(b_1,c_1)$, $(b_2,c_1)$, $(b_3,c_1)$, and  $(b_3,c_2)$.
The number of answers to $t$ is the  the sum of all multiplicities in the root. In our case it is  $4$. To sample an answer, we first sample $a \in R_{\{x\}}$ with weights given by the multiplicities. In our case we choose $a_1$ with probability 1. Then we sample $\big((a,m), (a,n)\big) \in R_{\{x,y\}} \times R_{\{x,z\}}$ such that $m\cdot n \geq 3$ with weights given by the product of the multiplicities of $(a,m)$ in $R'_{\{x,y\}}$ and  $(a,n)$ in $R'_{\{x,z\}}$. In our case, $\big((a_1,1),(a_1,3)\big)$ is chosen with probability $\frac{1}{2}$, and both $\big((a_1,2),(a_1,3)\big)$ and  $\big((a_1,2),(a_1,2)\big)$ with probability $\frac{1}{4}$. 
Finally, we sample $(a,b,m)$ and $(a,c,n)$ uniformly among relevant triples in $R_{\{x,y\}}$ and $R_{\{x,z\}}$, respectively, and we return $(a,b,c)$. In our case, for $\big((a_1,1),(a_1,3)\big)$ we choose either $\big((a_1,b_1,1),(a_1,c_1,3)\big)$ or $\big((a_1,b_2,1),(a_1,c_1,3)\big)$ with probability $\frac{1}{2}$, and for $\big((a_1,2),(a_1,3)\big)$ and  $\big((a_1,2),(a_1,2)\big)$ there is only one choice; overall, each answer is returned with probability $\frac{1}{4}$.
\end{example} 

\begin{algorithm}[t]
	\caption{Record sets $R_u$ for nodes $u \in U$ of decomposition $T$}
	\label{algo:records}
	\begin{algorithmic}[1]
\Loop { through nodes $u\in U$ of $T$ bottom-up}
	\If {$u$ is a $U$-leaf of $T$}  
	    \State $R_u \leftarrow  \big( q_u(D) \times \{1\} \big) \bowtie p_u(D, \fvar(p_u)\cap\fvar(t), c) $
	\EndIf
    \If {$u$ is a project node of $T$ with child $v\in U$} 
        \State $R_u \leftarrow  \gamma_{\mathit{witnessCount},\,X_u;\, \mathrm{sum}(\mathit{multiplicity})}\big(R_v\big)$
    \EndIf
    \If {$u$ is a join node of $T$ with children $v_1,v_2\in U$}  
        \State $R_{v_1,v_2,u}\leftarrow R_{v_1} \stackrel{c}{\bowtie} R_{v_2} \ltimes \check q_u(D)$
        \State $R'_{v_1,v_2, u} \leftarrow R_{v_1,v_2,u} \ltimes \check p_u(D) \uplus R_{v_1,v_2,u} \triangleright \check p_u(D)$
        \State $R_u \leftarrow \gamma_{\mathit{witnessCount},\,X_u;\, \mathrm{sum}(\mathit{multiplicity})}\big(R'_{v_1,v_2, u}\big)$
    \EndIf
    \If {$u$ is the root of $T$}  
         $R_u \gets \sigma_{a \leq \mathit{witnessCount} \leq b} \big(R_u\big)$
    \EndIf
\EndLoop
\end{algorithmic}
\end{algorithm}

\begin{theoremrep}
\label{thm:pseudopoly}
For TQs of free-connex tree-width $d$, over databases of size $n$, one can: count answers in time $\tilde O(n^{d})$; enumerate answers with constant delay after $\tilde O(n^{d})$ preprocessing; and sample answers uniformly at random in constant time after $\tilde O(n^{d})$ preprocessing.
 Assuming $d$ is constant, the combined complexity of each of  these algorithms is pseudopolynomial.
\end{theoremrep}
Compared to~\cite{BaganDG07,DurandM15}, this theorem provides the same complexity guarantees as for counting answers and enumerating answers for \emph{conjunctive queries}, but we are able to generalize these to \emph{threshold queries} (and add sampling). This generalization comes at the small cost of dependence on the value of the threshold. The results in \cite{BaganDG07,DurandM15} can be strengthened to more refined measures such as \emph{hypertreewidth}, but we believe that this is also true here.
%
\begin{proofsketch}
Consider a database $D$ and a threshold query 
\[ t(\bar x) = q(\bar x) \land \exists^{a,b} \, \bar y\: p(\bar x, \bar y)\,.\] 
We write $c$ for the threshold up to which we will be counting witnesses: if $b<\infty$, we let $c=b+1$; if $b=\infty$, we let $c=a$.
Let $T$ be a nice free-connex tree decomposition of $t$, of width $d$, and let $U$ be the maximal free-connex set in $T$.

Being a tree decomposition of $r(\bar x, \bar y) = q(\bar x) \land p(\bar x, \bar y)$, up to dropping unused variables, $T$ is a tree decomposition of both $p(\bar x)$ and $q(\bar x, \bar y)$. Let $q_u$ and $p_u$ be the corresponding subqueries associated to node $u$.
With each node $u \in U$ we associate the set $R_u$ of records that will provide information necessary to enumerate, count, and sample answers to $t$. 
A \emph{record for $u\in U$} is a triple $(\eta, n, k)$ consisting of a binding $\eta : X_u \to \adom(D)$, a \emph{multiplicity} $n>0$, and a \emph{witness count} $k \in \{0,1, \dots, c\}$ such that there are exactly $n$ extensions  $\eta'$ of $\eta$ to $\fvar(q_u)\cup \big(\fvar(p_u) \cap \fvar(t)\big)$ such that  $\pi_{\fvar(q_u)}(\eta') \in q_u(D)$ and $k = \min\big(c, |p_u(D, \eta')|\big)$.
We can interpret a record $(\eta, n, k)$ as a binding extending $\eta$ to two fresh variables, $\mathit{multiplicity}$ and $\mathit{witnessCount}$, with values $n$ and $k$. 

By Lemma~\ref{lem:safe}, each node in $u$ either is a $U$-leaf or has all its children in $U$. Consequently, we can compute $R_u$ for $u\in U$ bottom-up, as in  Algorithm~\ref{algo:records}. 
In $U$-leaves, we use Theorem~\ref{thm:answers_upto} to get a solution $p_u\big(D,\fvar(p_u)\cap\fvar(t), c\big)$ to the problem of counting answers to $p_u$ grouped by $\fvar(p_u)\cap\fvar(t)$ up to threshold $c$ over $D$;
the set $q_u(D)$ of all answers to $q_u$ over $D$ is computed as explained in Section~\ref{ssec:decompositions}. 
Higher up the tree, we compute $R_u$ based on the values obtained for the children of $u$, by means of standard relational operators with multiset semantics, including grouping ($\gamma$) and antijoin ($\triangleright$), as well as  the \emph{$c$-join} $R_{v_1} \stackrel{c}{\bowtie} R_{v_2}$  defined as the multiset of records $\big(\eta_1 \bowtie \eta_2, n_1 \cdot n_2, \min(c, k_1\cdot k_2)\big)$
such that $(\eta_1, n_1, k_1) \in R_{v_1}$, $(\eta_2, n_2, k_2) \in R_{v_2}$, and $\eta_1$ is compatible with $\eta_2$. Each $R_u$ is computed in $\tilde O(c\cdot |D|^d)$. With all $R_u$ at hand, we can enumerate, count, and sample answers to $t$ as in Example~\ref{ex:pseudopoly}. 
\end{proofsketch}
\begin{proof}
Consider a database $D$ and a threshold query 
\[ t(\bar x) = q(\bar x) \land \exists^{a,b} \, \bar y\: p(\bar x, \bar y)\,.\] 
We write $c$ for the threshold up to which we will be counting witnesses: if $b<\infty$, we let $c=b+1$; if $b=\infty$, we let $c=a$.
Let $T$ be a nice free-connex tree decomposition of $t$, of width $d$, and let $U$ be the maximal free-connex set in $T$.
%

Being a tree decomposition of $r(\bar x, \bar y) = q(\bar x) \land p(\bar x, \bar y)$, up to dropping unused variables, $T$ is a tree decomposition of both $p(\bar x)$ and $q(\bar x, \bar y)$. Let $q_u$ and $p_u$ be the corresponding subqueries associated to node $u$. It follows that $r_u = q_u\land p_u$.
By Lemma~\ref{lem:safe}, we have $X_u \subseteq \fvar(r_u) = \fvar(q_u)\cup \fvar(p_u)$. If $u \in U$ then additionally $X_u \subseteq \fvar(t)$ and $\fvar (q_u) \subseteq \fvar (q) = \fvar(t)$, and consequently
$ X_u \subseteq \fvar(q_u) \cup \big (\fvar(p_u) \cap \fvar(t))$.

With each node $u \in U$ we associate the set $R_u$ of records that will provide information necessary to enumerate, count, and sample answers to $t$. A \emph{record for $u\in U$} is a triple $(\eta, n, k)$ consisting of a binding $\eta : X_u \to \adom(D)$, a \emph{multiplicity} $n>0$, and a \emph{witness count} $k \in \{0,1, \dots, c\}$ such that there are exactly $n$ extensions  $\eta'$ of $\eta$ to $\fvar(q_u)\cup \big(\fvar(p_u) \cap \fvar(t)\big)$ such that  $\pi_{\fvar(q_u)}(\eta') \in q_u(D)$ and
$k = \min\big(c, |p_u(D, \eta')|\big)$.
We can interpret a record $(\eta, n, k)$ as a binding extending $\eta$ to two fresh variables, $\mathit{multiplicity}$ and $\mathit{witnessCount}$, with values $n$ and $k$.

By Lemma~\ref{lem:safe}, each node in $u$ either is a $U$-leaf or has all its children in $U$. Consequently, we can compute $R_u$ for $u\in U$ bottom-up. Consider Algorithm~\ref{algo:records}. Let $u$ be a $U$-leaf. From the definition of $U$ it follows that  $X_u = \fvar(q_u) \cup  \big(\fvar(p_u) \cap \fvar(t)\big)$. Let \[S_u = p_u\big(D,\fvar(p_u)\cap\fvar(t), c\big)\] be a solution to the problem of counting answers to $p_u$ grouped by $\fvar(p_u)\cap\fvar(t)$ up to threshold $c$ over $D$. By restricting all bags to $\var(p_u)$ and adding $\fvar(p_u) \cap \fvar(t)$ on top as the new root
we turn the subtree of $T$ rooted at $u$ into a $\fvar(p_u) \cap \fvar(t)$-rooted decomposition of $p_u$ of width at most $d$. Hence, we can use Theorem~\ref{thm:answers_upto} to compute $S_u$ in time $\tilde O(c\cdot|D|^d)$. The subtree of $T$ rooted at $u$ similarly induces an $\fvar(q_u)$-rooted decomposition of $q_u$ of width at most $d$, so we can compute the set $q_u(D)$ of all answers to $q_u$ over $D$ in time $\tilde O(|D|^d)$, as explained in Section~\ref{ssec:decompositions}. 
Then, we get $R_u$ as \[R_u = \big(q_u(D)\times \{1\}\big) \bowtie S_u\,;\] that is, the set of records $(\eta_1 \bowtie \eta_2, 1, k)$ such that $\eta_1 \in q_u(D)$, $(\eta_2, k) \in S_u$, and $\eta_1$ is compatible with $\eta_2$.
Note that all multiplicities are 1 because $X_u = \fvar(q_u)\cup \big(\fvar(p_u) \cap \fvar(t)\big)$, so only the trivial extension $\eta'=\eta$ is available. This step can be performed in $\tilde O\big(c\cdot |D|^d\big)$, as well. 


Suppose now that $u \in U$ is not a $U$-leaf. Then, all children of $u$ belong to $U$. Let us assume that their sets of records are already computed. 

Suppose that $u$ is a project node with child $v$. Then $R_u$ is obtained from $R_v$ by grouping records in $R_v$ by the projection to $X_u$ of the binding and by the witness count, and returning for each group the restricted binding, the sum of the multiplicities over the whole group, and the witness count. 
Because $|R_v| \leq (c+1)\cdot |D|^d$, this step can also be done in time $\tilde O\big(c\cdot |D|^d\big)$.

Suppose that $u$ is a join node with children $v_1$ and $v_2$.  First, we compute the \emph{$c$-join} \[R_{v_1,v_2}= R_{v_1} \stackrel{c}{\bowtie} R_{v_2}\] of  $R_{v_1}$ and $R_{v_2}$, defined as the multiset of records
\[\big(\eta_1 \bowtie \eta_2, n_1 \cdot n_2, \min(c, k_1\cdot k_2)\big)\]
such that $(\eta_1, n_1, k_1) \in R_{v_1}$, $(\eta_2, n_2, k_2) \in R_{v_2}$, and $\eta_1$ is compatible with $\eta_2$.
Next, we compute the \emph{semi-join} \[R_{v_1,v_2, u} = R_{v_1,v_2} \ltimes \check q_u(D)\] 
of $R_{v_1,v_2}$ with $\check q_u(D)$, defined as the multiset containing all occurrences of records $(\eta, n, k) \in  R_{v_1,v_2}$ such that $\pi_{\fvar(\check q_u)}(\eta) \in \check q_u(D)$.
Then, we compute the multiset 
\[R'_{v_1,v_2, u} = R_{v_1,v_2,u} \ltimes \check p_u(D) \uplus R_{v_1,v_2,u} \triangleright \check p_u(D)\] 
obtained as the multiset union of $R_{v_1,v_2,u} \ltimes \check p_u(D)$ and the \emph{anti-join} $R_{v_1,v_2,u} \triangleright \check p_u(D)$ of $R_{v_1,v_2,u}$ with $\check p_u(D)$, defined as the multiset containing one occurrence of $(\eta, n, 0)$ for each occurrence of record $(\eta, n, k) \in  R_{v_1,v_2,u}$ such that $\pi_{\fvar(\check p_u)}(\eta) \notin \check p_u(D)$. 
Finally, to compute $R_u$, we group the records $(\eta, n, k)$ in $R'_{v_1,v_2,u}$ by the binding $\eta$ and by the witness count $k$, and for each group we return the binding $\eta$, the sum of the multiplicities $n$ over the whole group, and the witness count $k$. Each of these operations can be performed in time $\tilde O(c\cdot |D|^d)$.

In the root we apply an additional filter to the computed records: we only keep records $(\eta, n,k)$ such that  $a \leq k \leq b$.

Apart from the records themselves we will need an index structure allowing quick access to relevant records in the children of each node. 
In a project node, for each stored record $(\eta, n, k)$ we need to be able to iterate over all records  $(\eta', n', k')$ from the child of $u$ such that  $\pi_{X_u}(\eta') = \eta$ and $k'=k$. 
In a join node $u$, similarly, for each stored record $(\eta, n, k)$ we need to be able to iterate over all pairs of records $(\eta_1, n_1, k_1)$ and $(\eta_2, n_2, k_2)$ from the respective children of $u$ that contribute to the multiplicity $n$. 
An appropriate index structure can be computed alongside the sets $R_u$. 
One way to do this is as follows. For a child $v$ of a project node $u$ we  store records $(\eta', n', k')$ of $R_v$ grouped by $\pi_{X_{u}}(\eta')$ and by $k'$; each record $(\eta,n,k)$ in $R_u$ is then equipped with a link to the group corresponding to $\eta$ and $k$. 
For children $v_1$ and $v_2$ of a join node $u$ we store the records of $R_{v_1}$ and $R_{v_2}$ grouped by the bindings; each record $(\eta,n,k)$ in $R_u$ is then equipped with a link to the group of $R_{v_i}$ that corresponds to $\pi_{X_{v_i}}(\eta)$ for $i=1, 2$. 
The pairs of records relevant to $(\eta,n,k)$ can be then read off the linked groups of $R_{v_1}$ and $R_{v_2}$: not all combinations of records from these groups are relevant, but skipping irrelevant ones involves only constant delay because both groups contain at most $c+1$ records.

Counting answers to $t$ based on the computed data structure is straightforward: the number of answers is the sum of all multiplicities of records stored in the root.

Enumeration can be done by iterating over stored records in a hierarchical fashion. That is, in each node the algorithm iterates over stored records and for each such record it iterates over relevant combinations of records in the children. When there is no next record to move to, the algorithm advances the iteration in the parent and updates the state of the iteration process in the affected descendants of the parent. The cost of each step is proportional to the size of the decomposition; that is, linear in the size of the query. 

Sampling the space of answers is similar to enumeration, but instead of iterating through the records, we choose them at random. Like in the case of enumeration we always ensure that records chosen in the children are compatible with the one chosen in their parent. Uniformity is guaranteed by using the probability distribution induced by the counts for records and combinations of records. In the root, we choose record $(\eta,n,k)$ with the probability $\frac{n}{m}$, where $m$ is the number of all answers to $t$. Suppose we have chosen a record $(\eta,n,k)$ in a project node $u$. Let \[(\eta_1, n_1, k_1), \dots,(\eta_\ell, n_\ell, k_\ell)\] be the relevant records in the child of $u$. Then, we should choose record $(\eta_i, n_i, k_i)$ with the probability $\frac{n_i}{n}$. Similarly, suppose we have chosen $(\eta,n,k)$ in a join node $u$. Let \[\big((\eta_1, n_1, k_1), (\eta'_1, n'_1, k'_1)\big), \dots, \big((\eta_\ell, n_\ell, k_\ell), (\eta'_\ell, n'_\ell, k'_\ell)\big) \]
be the relevant pairs of records from the children of $u$. Then, pair $\big((\eta_i, n_i, k_i), (\eta'_i, n'_i, k'_i)\big)$ should be chosen with probability $\frac{n_i\cdot n'_i}{n}$.
Using the \emph{alias method} \cite{Vose91,Walker77} we can precompute in time $O(n^d)$ an auxiliary data structure that allows performing the above random choices in constant time. 
\end{proof}



\section{Threshold Queries in the Wild}
\label{sec:use-cases}

In this section we give evidence that threshold queries are indeed common and useful in practice. Furthermore, we present an experimental evaluation of our algorithm. 

\subsection{Quantitative Study on Query Logs}

We first present some analytical results on large-scale real-world query logs from Wikidata's SPARQL query 
service \cite{WikiData-QueryService}.
Our study considers a corpus of more than 560M 
queries. (Previous work has considered a subset in the order of 200M queries \cite{bonifati2019navigating}.) 
These logs contain a massive amount of real-life queries, which are classified into 
(1) \emph{robotic} (high-volume, single-source bursts) and \emph{organic} (human-in-the-loop)~\cite{MalyshevKGGB18}.
Furthermore, the logs distinguish between successful (``OK'') and timeout (``TO'') requests.



\begin{table*}[htb]
\resizebox{!}{12mm}{
\begin{tabular}{@{}l@{\hspace{9mm}}c@{\hspace{9mm}}rr@{\hspace{9mm}}rr@{\hspace{9mm}}rr@{\hspace{9mm}}rr@{}}
    \toprule
    & \emph{Query Class} & \multicolumn{2}{c}{\emph{All}} & \multicolumn{2}{c}{\emph{Unique}} & \multicolumn{2}{c}{\emph{All ($\geq 2$)}} & \multicolumn{2}{c}{\emph{Unique ($\geq 2$)}}  \\
    \midrule
    & All Queries & 563,066,025 & 100.0\% & 74,060,492 & 100.0\% & 254,802,446 & 100.0\% & 30,502,206 & 100.0\%\\
    & CRPQ & 254,241,512 & 45.2\% & 41,778,884 & 56.4\% & 88,390,479 & 34.7\% & 3,346,280 & 11.0\%\\
    \midrule
    With \emph{Limit} & CRPQ & 15,225,933 & 6.0\% & 1,896,811 & 4.5\% & 13,205,975 & 14.9\% & 1,514,221 & 45.3\%\\
    With \emph{Limit}, no \emph{Order} & CRPQ & 15,214,527 & 6.0\% & 1,894,879 & 4.5\% & 13,195,363 & 14.9\% & 1,512,629 & 45.2\%\\
\bottomrule
\end{tabular}}
\smallskip
  \caption{Statistics of Wikidata query logs.\label{tab:querylogs} Percentages in the top half are relative to the total number of queries. Percentages in the bottom half are relative to the number of CRPQs. }
\end{table*}

\paragraph{Occurrences of Threshold Queries} 
We first report on the usage of the keywords \limit and \orderby in the Wikidata logs, which contain $\sim$563M well-formed queries, among which $\sim$74M are unique (Table~\ref{tab:querylogs}). Since our algorithms apply to \crpqs, we focus on those. As can be seen in the table, these still constitute 45.2\% of the queries and 56.4\% of the unique ones. In the remainder of this part, whenever we write a percentage as X\% (Y\%), then X refers to \emph{all} and Y to the \emph{unique} queries i.e., the set of queries after removing duplicates.

If we simply investigate how many \crpqs use \limit (columns \emph{All} and \emph{Unique}), these numbers are not so spectacular. Indeed, around 6\% (4.5\%) of the \crpqs use the \limit operator. What is remarkable though, is that almost all these queries that use \limit, \emph{do not} use an \orderby operation (bottom line of Table~\ref{tab:querylogs}).

By looking at the data more closely, we discovered that many of these queries are rather trivial in the sense that they only use one atom (or, equivalently, just one RDF triple pattern), which means that they do not even perform a join. For this reason, we decided to zoom in on the queries with \emph{at least two} atoms, see the columns \emph{All ($\geq\,2$)} and \emph{Unique ($\geq\,2$)}. It turns out that the numbers change significantly: around 14.9\% (45.3\%) of the \crpqs with at least two triples use \limit, which is  a significant amount. Again, we see that almost all the queries that use \limit, do not use \orderby. 

This is remarkable, because a commonly held belief is that \limit is most often used in combination with \orderby, i.e., as a means to express \emph{top-$k$ queries}. But, in this major real-world query log, this is not the case. Indeed, almost all non-trivial queries using \limit are threshold queries, seeking to return just an arbitrary \emph{unranked} sample of results. In fact, we have run the same analysis on a broader class of queries (\crpqf, which extend \crpq with unary filter conditions) and the percentages were very similar.




\begin{figure}[t]
\centering

\begin{tikzpicture}
\small
\tikzset{lines/.style={draw=white},}
\pie[
color={
  sred!60!white, 
  sviolet!60!white, 
  sblue!60!white, 
  sgreen!60!white,
  syellow!60!white, 
},
radius=1.2,
explode=0.0125,
sum=auto, 
after number={\scriptsize \%},
every only number node/.style={text=sbase03},style={lines}]
{
    41.4/,
    20.2/,
    24.8/,
    7.4/,
    6.3/
}
\begin{scope}[xshift=2cm,yshift=0.8cm,yscale=1.3]
\node[fill=syellow!60!white] at (0,0) {}; \node[right] at (0.125,0) {$\geq\,$10000};
\node[fill=sgreen!60!white] at (0,-0.35) {}; \node[right] at (0.125,-0.35) {1000 -- 9999};
\node[fill=sblue!60!white] at (0,-0.7) {}; \node[right] at (0.125,-0.7) {100 -- 999};
\node[fill=sviolet!60!white] at (0,-1.05) {}; \node[right] at (0.125,-1.05) {10 -- 99};
\node[fill=sred!60!white] at (0,-1.4) {}; \node[right] at (0.125,-1.4) {1 -- 9};
\end{scope}

\pie[pos={5.4,0},
color={
  sred!60!white, 
  sviolet!60!white, 
  sgreen!60!white,
  syellow!60!white, 
},
    radius=1.2,
    sum=auto, 
    explode=0.0125,
    after number={\scriptsize \%},
    every only number node/.style={text=base03},style={lines}]
{
63.4/,
12.0/,
9.2/,
15.3/
}

\node at (0,-1.5) {organic};
\node at (2.75,-1.5) {Threshold value};
\node at (5.5,-1.5) {robotic};

\end{tikzpicture}
\vspace{-3mm}
\caption{Threshold value occurrence ratio in \emph{organic} and \emph{robotic} CRPQ logs}
\label{all_limits_percentage}
\end{figure}


\paragraph{\limit Values} 
\newstuff{We now investigate the \emph{values} of the thresholds of queries in the logs. To this end, we considered the subset of CRPQs in the raw logs that use the keyword LIMIT ($\sim$15.2M queries, including duplicates). To gain deeper insight, we break down the logs into robotic ($\sim$15.2M) and organic ($\sim$44k) queries.
Fig.
~\ref{all_limits_percentage} shows the relative shares of threshold values with varying numbers of digits. The figure shows that threshold values between 1 and 9 are the most common. Still, in both organic and robotic queries, we see that large threshold values ($\geq\,$10K) are not uncommon. Since robotic logs can have large bursts of similar queries, we see that the distribution is not as smooth as for organic logs. For instance, only 0.08\% of queries in the robotic logs have three-digit limit values (depicted in blue), whereas there are much more (9.2\%) with four-digit values (depicted in green). The largest value we found in robotic logs had 9 digits. In organic logs, however, we found three limit values containing 11, 14, and 17 digits, respectively.}

\begin{toappendix}
\begin{figure}
\begin{tikzpicture}
\begin{axis}[ybar,
yscale = 0.5,
enlargelimits=0.15,
ylabel=Limit $k$,
bar width=3pt,
legend style={
  at={(0.5,-0.7)},
  anchor=north,
  legend columns=-1
},
symbolic x coords={1-10,11-20,21-30,31-40,41-50,51-60,61-70,71-80,81-90,91-100},
x tick label style={rotate=45,anchor=east},
xtick=data,
]
\addplot[fill=sred!60!white,draw=sred!60!white] coordinates{(1-10,3221770) (11-20,1700422) (21-30,2558) (31-40,2955) (41-50,166224) (51-60,64) (61-70,92) (71-80,363) (81-90,291) (91-100,506810)}; 
\addplot[fill=sviolet!60!white,draw=sviolet!60!white] coordinates{(1-10,6436871) (11-20,4227418) (21-30,4219) (31-40,3846) (41-50,188107) (51-60,117) (61-70,129) (71-80,414) (81-90,334) (91-100,631116)}; 
\legend{\crpq log, \crpqf  log}
\end{axis}
\end{tikzpicture}
\vspace{-3mm}
\caption{Distribution of threshold values (1-100 interval) in mixed (organic and robotic) logs}
\label{fig:CRPQ-CRPQf-small-limits}
\end{figure}
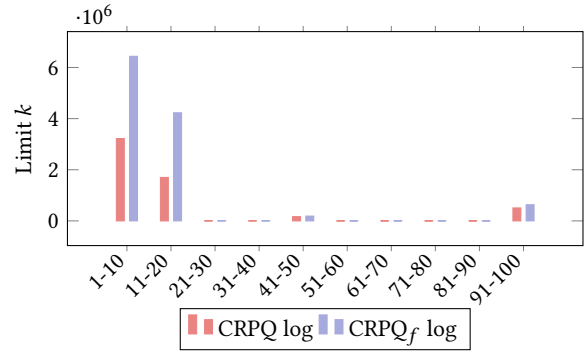

\paragraph{\limit Values} 
In addition to the analysis on \limit values in the body, we also investigated the distribution of values below 100. The result can be found in Figure~\ref{fig:CRPQ-CRPQf-small-limits}. Indeed, when focusing on these threshold values $<$ 100, we notice that in both \crpq and \crpqfs the most occurrences correspond to values $< 10$, followed by those $< 20$, by those ranging from 91--100, with few occurrences for the interval 41--50 and outliers.  
\end{toappendix}

\paragraph{Conclusion of Quantitative Study} Threshold queries are indeed quite common, e.g., in querying knowledge bases such as Wikidata. Since the actual values of the thresholds are typically small, our empirical study confirms the utility in practice of our pseudopolynomial algorithm (Theorem~\ref{thm:pseudopoly}) that, in order to evaluate queries with a threshold $k$, solely needs to maintain up to $k+1$ intermediate results per each candidate answer tuple. This is in contrast with traditional query plans where the number of intermediate results per candidate answer tuple is determined by the input data and therefore potentially orders of magnitude larger.

\begin{toappendix}
\paragraph{SPARQL Terms} 
In order to measure the size of the threshold queries in the logs (\crpqs and \crpqfs as shown in Table \ref{tab:querylogs}), we have counted the total number of triple patterns contained in the clauses \texttt{SELECT}, \texttt{ASK}, and \texttt{CONSTRUCT} of those queries. Triple patterns are similar to RDF triples, 
except that each of the subject, predicate, and object can be a variable. The results show that most threshold queries use two triple patterns (54,59\%), and that queries with one or three triple patterns are also quite frequent (15,41\% and 29,80\%, respectively) . 

Overall, we can observe that 99,8\% of the Wikidata queries in the \crpqf corpus use at most three triple patterns. This trend is similar to query sizes of the entire corpus, as studied in previous work \cite{bonifati2019navigating}. 
We observed that organic queries are more diverse in terms of sizes with respect to robotic queries. 

We also analyzed the individual components in triple patterns, such as blank nodes, SPARQL variables, IRIs, and literals. 
Table \ref{tab:tpstats} shows the results for the entire corpus including OK and TO queries. 
Blank nodes are existential variables; that is, they state the existence of an entity without identifying it, while SPARQL variables can be seen as universal variables. To emphasize  the relationship between variables and blank nodes, we compute the $rvb$ ratio 
between variables and, respectively, blank nodes and variables. 
Similarly, we record the $ril$ ratio 
between IRIs and, respectively, IRIs and literals (both of which  can be regarded as constants). 

The results, including $rvb$ and $ril$ ratios, show that the most common terms are SPARQL variables and IRIs. This is consistent with the fact that IRI references are the basis for Linked Data and that variables are at the core of SPARQL. 
We note that blank nodes are the least frequent and, when used, occur only once per query in 99.96\% of cases. 
Conversely, IRIs, which provide a mechanism for globally identifying resources, are much more common.

\begin{table}[]
\begin{tabular}{@{}lll@{}}
\toprule
            & \emph{CRPQ}     & \emph{CRPQ$_f$}    \\
\midrule
blank nodes & 339,219   & 3,390,834  \\
 --- blank nodes in TO & 20 & 93 \\
variables  & 9,582,508  & 24,359,265 \\
 --- variables in TO & 1398 & 6849 \\
\midrule
$rvb$       & 96.58\%  & 87.78\%  \\
$rvb_{TO}$  & 98.59\%  & 98.66\%  \\
\midrule
IRIs        & 18,285,340 & 22,444,709 \\
--- properties & 1,078,860 & 1,204,279 \\ 
--- items  & 2,535,292 & 5,523,424    \\ 
--- IRIs in TO & 1,373 & 7,147 \\
literals    & 369,825   & 4,371,653  \\
--- literals in TO & 34 & 2,720 \\
\midrule
$ril$       & 98.02\%  & 83.7\%   \\
$ril_{TO}$  & 97.58\%  & 72.43\%   \\
\bottomrule
\end{tabular}
\caption{Statistics of SPARQL terms for (OK+TO) and TO only queries.\label{tab:tpstats}}
\end{table}

We also observed that timeout queries usually contained more variables and blank nodes. This relationship does not apply to almost IRIs and literals. 
\end{toappendix}

\subsection{Qualitative Study}

To demonstrate further the usefulness of threshold queries in practice across diverse contemporary domains, we also performed a qualitative study on two real-world graph datasets.

\paragraph{Covid-19 Dataset}
The Covid-19 Knowledge Graph \cite{CovidGraph} 
is a continuously evolving dataset, with more than 10M nodes and 25M edges, obtained by integrating various data sources, including gene expression repositories (e.g., the Genotype Tissue Expression (GTEx) and the COVID-19 Disease Map genes), as well as article collections from different scientific domains (ArXiv, BioRxiv, MedRxiv, PubMed, and PubMed Central). 
The inferred schema of this graph exhibits more than 60 distinct node labels and more than 70 distinct edge labels \cite{LbathBH21}. 
Such typing information is, however, not sufficient to express the domain-specific constraints that can be found in these real-life graph datasets. 
Non-trivial constraints expressible with threshold queries can be naturally crafted in order to complement the schema, as we showcase in our study. 

\paragraph{Wikidata Dataset}
Wikidata is a collaborative knowledge base launched in 2012 and hosted by the Wikimedia Foundation~\cite{vrandecic2012}. 
By the efforts of thousands of volunteers, the project has produced a large, open knowledge base with numerous applications.
Wikidata can be seen as a graph database which has a SPARQL endpoint that lets users and APIs formulate queries on the its massive knowledge graph. 
The query logs collected along the years on this endpoint \cite{krotzsch2018practical} constitute a useful resource for the qualitative analysis of threshold queries.  

\begin{figure}[t!]
\tikzset{ex/.style={inner sep=2.5pt}}
\tikzset{fr/.style={draw,inner sep=2.5pt}}
\tikzset{tl/.style={draw,rounded corners,inner sep=2.5pt}}
\tikzset{output/.style={draw,rounded corners,double,inner sep=2.5pt}}

\centering
\begin{tikzpicture}[>=stealth']
\footnotesize

\begin{scope}[xshift=.25cm,yshift=0cm]
\node at (-1.25,0) {\normalsize\bf TQ1};
\node[output] (name) at (-.345,0) {\it ?name}; 
\node[tl] (person) at (1,0) {\it ?person};
\node at (1,-.35) {$\exists\mathbin{>}\!1000$};

\draw (name) edge[<-] (person);
\end{scope}

\begin{scope}[xshift=3.5cm,yshift=-0cm]
\node at (-1,0) {\normalsize\bf TQ2};
\node[output] (protein) at (0,0) {\it p:Protein}; 
\node[tl] (term) at (2.5,0) {\it t:GOTerm};
\node at (2.5,-.35) {$\exists\mathbin{>}43917$};
\draw (protein) edge[->,decorate,decoration={snake,amplitude=.2mm}] node[above]{\scriptsize \it maps$^*$...} (term);
\end{scope}



\begin{scope}[xshift=.25cm,yshift=-1cm]
\node at (-1.25,0) {\normalsize\bf TQ3};
\node[output] (country) at (-0.2,0) {\it c:Country};
\node[output] (age) at ($(country)+(3,0)$) {\it a:AgeGroup};
\draw (country) edge[->] node(e)[tl,above,yshift=.5mm]{\;\scriptsize \it e\;} (age);
\node at ($(e)-(0,.4)$) {$\exists < 3$};
\end{scope}

\begin{scope}[xshift=2.8cm,yshift=-2cm]
\node at (-3.8,0) {\normalsize\bf TQ4};

\node[tl] (x3) at (0,.25) {$x_3$}; 
\node[output] (x1) at ($(x3)+(-3.0,-.25)$) {$x_1$}; 
\node[output] (x2) at ($(x1)+(.25,-1.5)$) {$x_2$}; 
\node[output] (x4) at ($(x3)+(3.0,-.25)$) {$x_4$}; 
\node[output] (x5) at ($(x4)+(-.25,-1.5)$) {$x_5$}; 
\node[output] (x6) at (0,-2) {$x_6$}; 
\node[output] (x7) at (-2,-1.875) {$x_7$}; 
\node[tl] (x8) at ($(x6)+(1.75,.125)$) {$x_8$};

\draw[bend angle=05] (x1) edge[->,bend right] node[below,sloped] {\it \scriptsize protein ID} (x2);
\draw[bend angle=05] (x1) edge[->,bend left] node[below,sloped,text centered,text width=3cm,pos=0.45] {\it \scriptsize molecular function $\mid$ \\cell component $\mid$ \\biological process} (x3);
\draw[bend angle=05] (x3) edge[->,bend left] node[below,sloped,text centered,text width=3cm,pos=0.55] {\it \scriptsize molecular function $\mid$ \\cell component $\mid$ \\biological process} (x4);
\draw[bend angle=05] (x4) edge[->,bend left] node[below,sloped] {\it \scriptsize instance of} (x5);
\draw (x3) edge[->] node[above,sloped,text centered,text width=2.25cm] {\it \scriptsize was derived from $/$ \\stated in} (x6);
\draw[bend angle=05] (x6) edge[->,bend left] node[below,sloped] {\it \scriptsize PubMed ID} (x7);
\draw[bend angle=05] (x6) edge[->,bend right] node[below,sloped] {\it \scriptsize published in} (x8);

\node at (3,-2.2) {{\tt LIMIT} 1000};
\end{scope}
\end{tikzpicture}
\caption{Structural diagrams of selected threshold queries}
\label{fig:threshold-queries}
\end{figure}

\paragraph{Working Method}
We have manually inspected the Covid-19 Knowledge Graph schema in search of constraints that can be validated with threshold queries. We have also found a number of threshold queries by sieving through a large sample of Wikidata query logs. We analyzed the structural properties of the collected threshold queries. Below we discuss our findings with the help of five selected threshold queries from the two datasets. 
The queries are depicted in Figure~\ref{fig:threshold-queries}. We focus on the structure of the underlying graph patters, omitting most of the labels. Constants are in quotes, variables are in rounded rectangles, and output variables have a double edge. Wavy edges represent path expressions in the original query. For readability, we sometimes change the labels of nodes in the presented queries.

\paragraph{Selected Queries}
The first example comes from the Wikidata logs and appears to be a \emph{data exploration} query.
\begin{enumerate}
  \itemsep0pt
\item[TQ1] \sl 
Return all given names with more than 1000 occurrences.
\end{enumerate}
\begin{lstlisting}
    SELECT ?name WHERE { ?person <given_name> ?name } 
    GROUP BY ?name  HAVING ( ( COUNT (*) > 1000 ) );
\end{lstlisting}


A structurally similar query can help detect \emph{integrity violations} in the Covid-19 Knowledge Graph. Namely, the latest release (June 2021) of the Gene Ontology (GO) contains $43917$ valid terms \cite{GOrelease, GOpaper}. Therefore, in the Covid-19 Knowledge Graph one can check whether a protein is suspicious if it exhibits more than this number of GO terms associated to it.
\begin{enumerate}
  \itemsep0pt
\item[TQ2] \sl Find each protein that has more than $43917$ associated gene ontology terms.
\end{enumerate}
A formulation of this query in Cypher-like syntax follows.\footnote{Note that Cypher does not currently support concatenation (\texttt{:A\,.\,:B}).}
\begin{lstlisting}
    MATCH (p:Protein)-[:MAPS* . :HAS_ASSOCIATION . 
                      (:IS_A*|PART_OF*)]->(t:GOTerm) 
    WITH p, COUNT(DISTINCT t) AS count_go
    WHERE count_go > 43917 RETURN p;
\end{lstlisting}
Notice the large threshold and the complex regular expression.

%
As another example, reporting coverage in the Covid-19 Knowledge Graph demands that for each age group, in each country, there should be at least three 
reports for the current number of Covid cases (one for females, one for males, and one for the total). Deviations can be identified with the following threshold query. 
\begin{enumerate}
  \itemsep0pt
\item[TQ3] \sl Find each country that does not have three reports for some age group. 
\end{enumerate}
This query can be expressed in Cypher-like syntax as follows. 
\begin{lstlisting} 
    MATCH (c:Country)-[e:CURRENT_FEMALE|CURRENT_MALE
                      |CURRENT_TOTAL]->(a:AgeGroup)
    WITH c, a, COUNT(type(e)) AS ecount 
    WHERE ecount < 3 RETURN c, a;
\end{lstlisting}
We point out that this query although acyclic is not free-connex. 

We also discovered that graph patterns in limit-$k$ queries can be quite large, as in the following query from the Wikidata logs.
\begin{enumerate}
  \itemsep0pt
\item[TQ4] \sl 
Return 1000 tuples containing names (x1) and UniProt IDs (x2) for genes with domains (x3)
referenced in “Science”, “Nature,” or “Cell” articles, as well as their indirect domain label (x4) and class (x5), the article name (x6), and its PubMed ID (x7).
\end{enumerate}

\paragraph{Conclusion of Qualitative Study} Our qualitative study discovered several interesting and realistic uses of threshold queries that we illustrated here. Connecting these to our algorithms, it is interesting to note that all these queries are acyclic, but only TQ1-TQ2 are free-connex (Fig.~\ref{fig:threshold-queries}). 
For instance, TQ4 has star size 3 (due to $x_3$, which has 3 neighboring nodes that propagate to the output) and free-connex tree-width 4. 
Evaluation of such queries with existing querying approaches may incur significant cost while our algorithms handle them very efficiently.
For instance, TQ4 can be evaluated in linear time by Algorithm~\ref{algo:grouped-answers-threshold} (up to logarithmic factors), while the approach via enumeration \cite{BaganDG07} requires at least cubic time.



\begin{toappendix}
Another potential purpose for threshold queries we identified is \emph{data monitoring}, exemplified by the following Boolean query from the Wikidata logs.  
\begin{enumerate}
  \itemsep0pt
\item[TQ5] \sl 
Are there at least 67 language versions of Wikipedia pages for the movie \emph{The Matrix}?
\end{enumerate}

\OMIT{Query TQ5 was \red{found in the SPARQL logs as}: \wim{This query is written in such a convoluted way that I think that it's better not in the body. The pattern in the figure is equivalent and much clearer.}
\begin{lstlisting}
    SELECT( ?var2 ) WHERE { 
      VALUES ( ?var2 ) { ( <The_Matrix> ) } 
      ?var2 <instance_of> <film> . 
      ?var3 <about> ?var2 . 
      ?var3 ( <is_Part_Of> / <belongs_to> ) "wikipedia" . 
    } GROUP BY ?var2 HAVING ( (  COUNT (*) >= 67 ) );
\end{lstlisting}}
\end{toappendix}

\newstuff{\subsection{Experimental Study}}

\newstuff{As a proof-of-concept, we implemented the algorithm in Theorem~\ref{thm:answers_upto} in SQL
and compared it with the optimizer of a popular DBMS engine, namely 
the PostgreSQL 13.4 optimizer.
Using SQL windowing functions, our implementation can mimic some important aspects of our query evaluation algorithm (like internal information passing up to a threshold), but we note that SQL does not allow to capture our algorithm precisely. 
As shown in the remainder, the results of this comparison already show the superiority of our algorithm for threshold queries 
compared to naive evaluation.}

\newstuff{All the experiments were executed on an Intel Core i7-4770K CPU @ 3.50GHz, 
16GB of RAM, and an SSD.
We used PostgreSQL 13.4 in Linux Mint 20.2 
and built our own micro-benchmark \cite{implementation} consisting of the following three types of query templates:
\begin{enumerate}[(q1)]
\item 
    $k$-\texttt{path} selects up to 10 pairs of nodes linked by a $k$-hop path;
\item 
    $k$-\texttt{neigh} selects all nodes with $\geq$ 10 $k$-hop neighbors;
\item 
    $k$-\texttt{conn} selects all pairs of nodes linked by $\geq$ 10 $k$-hop paths.
\end{enumerate}
Assuming that the database consists of a single binary relation $R$, then these queries for $k=2$ can be naturally formulated in SQL as:}
\begin{lstlisting}
  SELECT DISTINCT R1.A, R2.B FROM R AS R1, R AS R2 
  WHERE R1.B = R2.A LIMIT 10;
\end{lstlisting}

\begin{lstlisting}
  SELECT R1.A FROM R AS R1, R AS R2 WHERE R1.B = R2.A 
  GROUP BY R1.A HAVING COUNT(DISTINCT R2.B) >= 10;
\end{lstlisting}

\begin{lstlisting}
  SELECT X0, X2 FROM 
     (SELECT DISTINCT R1.A AS X0, R1.B AS X1, R2.B AS X2
     FROM R AS R1, R AS R2 WHERE R1.B = R2.A) AS SUB
  GROUP BY X0, X2 HAVING COUNT(*) >= 10;
\end{lstlisting}
\newstuff{In the following, we will refer to these formulations as the \emph{baseline} formulations of the queries.

We compared these queries with alternative formulations in SQL that mimic crucial aspects of our evaluation algorithm and that can be understood as follows. A simple decomposition for $k$-\texttt{path} is a tree with a single branch and one node per each joined copy of table $\texttt{R}$. For this decomposition and $k=2$ the algorithm in Theorem~\ref{thm:answers_upto} amounts to evaluating the following query:}
\begin{lstlisting}
  WITH
  J1 AS (SELECT DISTINCT A AS X1, B AS X2 FROM R),
  W1 AS (SELECT X1, X2, 
                ROW_NUMBER() OVER (PARTITION BY X1) AS RK
         FROM J1),
  S1 AS (SELECT X1, X2 FROM W1 WHERE RK <= 10),
  J2 AS (SELECT DISTINCT R.A AS X0, X2 
         FROM R, S1 WHERE R.B = S1.X1),
  W2 AS (SELECT X0, X2, 
                ROW_NUMBER() OVER (PARTITION BY X0) AS RK
         FROM J2),
  S2 AS (SELECT X0, X2 FROM W2 WHERE RK <= 10)
  SELECT X0, X2 FROM S2 LIMIT 10;
\end{lstlisting}
\newstuff{For $k$-\texttt{neigh} we can use the same decomposition and the corresponding query is the same except for the last line which is}
\OMIT{\begin{lstlisting}
WITH
J1 AS (SELECT DISTINCT A AS X1, B AS X2 FROM R),
W1 AS (SELECT X1, X2, 
              ROW_NUMBER() OVER (PARTITION BY X1) AS RK
       FROM J1),
S1 AS (SELECT X1, X2 FROM W1 WHERE RK <= 10),
J2 AS (SELECT DISTINCT R.A AS X0, X2 
       FROM R, S1 WHERE R.B = S1.X1),
W2 AS (SELECT X0, X2, 
              ROW_NUMBER() OVER (PARTITION BY X0) AS RK 
       FROM J2),
S2 AS (SELECT X0, X2 FROM W2 WHERE RK <= 10)
SELECT X0 FROM S1 GROUP BY X0 HAVING count(*)>=10;
\end{lstlisting}}
\begin{lstlisting}
  SELECT X0 FROM S1 GROUP BY X0 HAVING count(*)>=10;
\end{lstlisting}
\newstuff{For $k$-\texttt{conn} we can also use the same decomposition but the corresponding query is a bit different, as we need to collect the whole paths, not just their endpoints. We refer to these alternative implementations as the \emph{windowed} versions of the queries. Natural query plans for the windowed versions have worst-case complexity $\tilde O(k\cdot n)$ for $k$-\texttt{path} and $k$-\texttt{neigh}, and  $\tilde O(k\cdot n^2)$ for $k$-\texttt{conn}.}

\begin{table*}[t]
\resizebox{!}{12.4mm}{
\newstuff{
\begin{tabular}{@{}lrrrrrrrrrr@{}}
    \toprule
    $k$ & 1 & 2 & 3 & 4 & 5 & 6 & 7 & 8 & 9 & 10\\
    \midrule
    q1/b & {\bf 39} & 277 & 5,157 & 103,449 & T/O & T/O & T/O & T/O & T/O & T/O \\
    q1/w & {\bf 39} &  {\bf 58} &   {\bf  69} &    {\bf   88} &  {\bf 97} & {\bf 115} & {\bf 132} & {\bf 148} & {\bf 162} & {\bf 176} \\
    \midrule
    q2/b & 50 & 349 & 5,742 & 134,970 & T/O & T/O & T/O & T/O & T/O & T/O \\
    q2/w & {\bf 44} &  {\bf 63} &    {\bf 74} &     {\bf  93} & {\bf 104} & {\bf 119} & {\bf 137} & {\bf 150} & {\bf 167} & {\bf 180} \\
    \midrule
    q3/b &  {\bf 94} & {\bf 652} & 11,568 & 267,176 & T/O & T/O & T/O & T/O & T/O & T/O  \\
    q3/w & 146 & 690 &  {\bf 2,695} &  {\bf 5,266} & {\bf 10,679} & {\bf 18,400} & {\bf 31,020} & {\bf 48,942} & {\bf 74,795} & {\bf 103,054}\\
\bottomrule
\end{tabular}}
} \hfill \resizebox{!}{12.4mm}{
\newstuff{
\begin{tabular}{@{}rrrrrrrrrr@{}}
    \toprule
    1 & 2 & 3 & 4 & 5 & 6 & 7 & 8 & 9 & 10\\
    \midrule
    {\bf 74} & 549 & 3,364 & 19,161 & 106,741 & 601,279 &   T/O &   T/O &   T/O &   T/O \\
    80 & {\bf 196} &   {\bf 298} &    {\bf 404} &    {\bf  537} &     {\bf 665} & {\bf 1,260} & {\bf 1,478} & {\bf 1,547} & {\bf 1,675}  \\
    \midrule
    {\bf 138} & 1,296 & 11,207 & 90,199 & 644,860 &  T/O &  T/O &  T/O &  T/O &  T/O \\
    152 &   {\bf 282} &    {\bf 399} &    {\bf 506} &    {\bf  579} &  {\bf 653} & {\bf 1,235} & {\bf 1,390} & {\bf 1,540 }& {\bf 1,667}\\
    \midrule
    {\bf 273} & {\bf 2,423} & {\bf 18,670} & 122,894 & T/O & T/O & T/O & T/O & T/O & T/O\\
    420 & 4,201 & {\bf 23,903} & {\bf 57,555} & {\bf 109,764} & {\bf 189,893} & {\bf 301,510} & {\bf 390,968} & {\bf 510,171} & {\bf 576,009}\\
    \bottomrule
\end{tabular}}
}
\smallskip
 \caption{\newstuff{
Experimental evaluation for the baseline (b) and windowed (w) versions of (q1--q3) on the IMDb database (left) and on Barab\'asi-Albert graphs (right) with $m_0 = 10$ and $n=3000$. Time is measured in \textbf{ms}. Better running time is depicted in bold.\label{tab:experiments}}}
\end{table*}

\newstuff{
\paragraph{Datasets} We considered two kinds of data sets:
\begin{enumerate}[(1)]
\item The real-world IMDb data set used in Join Order Benchmark~\cite{leis2018query}, which contains information about movies and related facts about actors, directors, production companies, etc. In our experiments, we focused on the \emph{movie\_link} relation, which has a graph-like structure, so that finding paths is meaningful.
\item A number of synthetic data sets, which are Barab\'asi-Albert graphs \cite{BarabasiAlbert} with varying parameters of $n$ (total number of nodes to add) and $m_0$ (the number of edges to attach from newly added nodes to existing nodes).\footnote{Barab\'asi-Albert graphs are a kind of preferential attachment graphs that are scale-free and model the structure of e.g.\ social networks or the internet.}
\end{enumerate}
We repeated each experiment several times and report the median. 

\paragraph{First Experiment}
We compared the running time of the baseline and windowed versions of (q1--q3) for varying values of $k$ on both the IMDb and synthetic data sets, both on fully indexed and non-indexed databases. Table~\ref{tab:experiments} shows the results for the non-indexed case. We see that our approach (windowed) outperforms the baseline with speedups up to three orders of magnitude, while the baseline times out (T/O) for higher values of $k$. The running times of the windowed versions reflect the good theoretical bounds, while the baseline clearly shows exponential dependence on $k$. For the indexed case, leveraging the DBMS's default indexes, the baseline ran only marginally faster ($\sim$5\%), remaining three orders of magnitude slower than our algorithm.
}


\begin{table}[t]
\setlength{\tabcolsep}{6.1pt}
 \resizebox{!}{11mm}{
\newstuff{
 \begin{tabular}{@{}lrrrrrrrrrr@{}}
 \toprule
 $m_0 \backslash n$ & 32 & 100 & 316 & 1k & 3.2k & 10k & 32k & 100k & 316k & 1M\\
 \midrule
 5 & 0.6 & 1.8 & 3.0 & 3.6 & 4.6 & 5.7& 6.7 & 7.7 & 8.7 & 10.0 \\
 10 & 1.4 & 6.8 & 13.8 & 22.8 & 29.7 & 37.1 & 40.0 & 73.5 & T/O & T/O\\
 15 & 2.1 & 16.3 & 46.6 & 66.2 & 90.5 & 120.4 & T/O & T/O & T/O & T/O\\
 20 & 1.1 & 35.4 & 96.0 & 149.1 & 192.8 & 404.3 & T/O & T/O & T/O & T/O\\
 25 & 0.6 & 55.2 & 184.7 & 272.5 & 351.7 & T/O & T/O & T/O & T/O & T/O \\   
 \bottomrule
 \end{tabular}
}}
\smallskip
\caption{\newstuff{Experimental evaluation of the speedup factor (the ratio of baseline to windowed) for q2 on Barab\'asi-Albert graphs for varying $n$ and $m_0$.}\label{tab:heatmap}}
\end{table}

\OMIT{
\begin{table}[t]
\setlength{\tabcolsep}{3pt}
 \resizebox{!}{18mm}{
\begin{tabular}{lrrrrr}
 \toprule
 $n \backslash m_0$ & 5 & 10 & 15 & 20 & 25 \\   
 \midrule
 32 &  0.6 & 1.4 & 2.1 & 1.1 & 0.6 \\
 100 & 1.8 & 6.8 & 16.3 & 35.4 & 55.2 \\
 316 & 3.0 & 13.8 & 46.6 & 96.0 & 184.7 \\
 1000 & 3.6 & 22.8 & 66.2 & 149.1 & 272.5 \\
 3162 & 4.6 & 29.7 & 90.5 & 192.8 & 351.7 \\
 10000& 5.7& 37.1 & 120.4 & 404.3 & T/O$_b$ \\
 31623 & 6.7 & 40.0 & T/O$_b$ & T/O$_b$ & T/O$_b$ \\
 100000 & 7.7 & 73.5 & T/O$_b$ & T/O$_b$ & T/O$_b$ \\
 316228 & 8.7 & T/O$_b$ & T/O$_b$ & T/O$_b$ & T/O$_b$\\
 1000000& 10.0 & T/O$_b$ & T/O$_b$ & T/O$_b$ & T/O$_b$ \\
 \bottomrule
 \end{tabular}
 \hspace{5pt}
 \begin{tabular}{rrrrr}
 \toprule
  5 & 10 & 15 & 20 & 25 \\   
 \midrule
0.6 & 1.4 & 1.2 & 0.8 & 0.6 \\
1.1 & 4.6 & 7.7 & 13.4 & 27.7 \\
1.5 & 6.4 & 14.7 & 25.3 & 40.7 \\
2.1 & 9.1 & 18.5 & 30.2 & 52.9 \\
3.4 & 13.6 & 29.4 & 56.7 & 85.9 \\
3.6 & 14.7 & 38.9 & 67.2 &  \\
3.3 & 15.1 & 35.6 & 83.1 &  \\
3.2 & 14.3 &  &  &  \\
3.4 & 16.5 &  &  &  \\
3.2 &  &  &  & 	\\ 
\bottomrule
 \end{tabular}
 }
\caption{Experimental evaluation of the speedup factor (measured as the ratio of baseline to windowed) for q2 (left) and q1 (right) on Barab\'asi-Albert graphs for varying $n$ and $m_0$.}
\end{table}
}

\newstuff{
\paragraph{Second Experiment}
In this second experiment, we wanted to assess the impact of the structure and size of the data on the runtimes of our algorithm. To this purpose, we have employed the Barab\'asi-Albert synthetic graphs with varying outdegree ($m_0$) and number of nodes ($n$) and considered query q2 with $k=3$. Table~\ref{tab:heatmap} shows the different speedups that the windowed approach offers, when compared to the baseline. We varied $m_0$ from 5 to 25, using increments of 5, and varied $n$ from 32 to 1M in a logarithmic scale with increment factor of $\sqrt{10} \approx 3.16$. 
In the table, we see that the speedup factor of the windowed approach increases by up to three orders of magnitude as the size of the data and out-degree $m_0$ increase. T/O means that the baseline approach timed out (> 30 minutes). For all entries in Table~\ref{tab:heatmap}, the windowed algorithm terminated under 15 minutes. These results show the robustness of our algorithm to variations of dataset size and outdegree as well as its superiority with respect to the baseline.}

\OMIT{
Several points: 
\begin{itemize}
\item T/O$_b$ means that the baseline times out, T/O$_w$ that the windowed version times out. No subscript: both time out. We could just use one kind of annotation, meaning "slower times out". Or no annotations at all and just say it in the text.

\item We do not show plots for q1 because the results are very similar (BTW, this should be verified). 

\item There are two reasons why our performance for q3 is not as impressive for q1 or q2. The first reason is that our algorithm for q3 runs in time $\tilde O(n^2)$ over graphs with $n$ vertices, while the one for q1 runs in time $\tilde O(n)$. This is why the windowed version of q3 times out for large graphs, while q2 terminates in under 15 minutes even for $n=1,000,000$ and $m_0=25$. The second reason is that the speedup depends on the fraction of partial results that is filtered out by our pruning techniques. Even in not so dense graphs, a node often has many $k$-hop neighbours, so keeping only up to 10 leads to huge savings for q1 and q2. On the other hand, the number of $k$-hop paths between any given endpoints need not be large, so the savings offered by keeping only up to 10 paths for each pair of endpoints may not balance out the cost of pruning. For example, consider Barab\'asi-Albert graphs with $n=10000$: for $m_0 = 10$ only 15\% of $3$-hop paths are pruned, but for $m_0 = 20$ the fraction grows to up to $51\%$. Savings are larger for longer paths: 71\% of $4$-hop paths are pruned already for$m_0=10$.
\end{itemize}
}

\section{Related Work}
\label{sec:related-work}


\paragraph{Top-$k$ and Any-$k$}

The potential of predefined thresholds to speed up query processing was first noticed by Carey and Kossmann \cite{Carey1997}, who explored ways of propagating thresholds down query plans, \newstuff{dubbed \emph{\limit pushing}}. This early study only considered applying the thresholds directly to subplans, which made joins a formidable obstacle. In contrast, we first group the answers to subqueries by variables determined based on the structure of the whole query, and then apply thresholds within groups; this way we can push thresholds down through multi-way joins, guided by a tree decomposition of the query. 
Most follow-up work concerns the ranked scenario, where the goal is to compute \emph{top-$k$} answers according to a specified preference order. The celebrated Threshold Algorithm \cite{Fagin2003} solves the top-$k$ selection problem: it operates on a single vertically-partitioned table, with each partition being managed by a different external service that only knows the scores of base tuples in its own partition, and produces $k$ tuples with the highest score while minimizing the number of reads. There are also multiple approaches to the more general top-$k$ join problem.
J*~\cite{natsev2001supporting} is based on the A* search algorithm: it maintains a priority queue of partial and complete join combinations ordered by the upper bounds of their scores. 
Rank-Join~\cite{ilyas2004supporting} maintains scores for complete join combinations only, and stops when new combinations cannot improve the current top-$k$.  
LARA-J*~\cite{mamoulis2007efficient} offers improved handling of multiway joins.  FRPA~\cite{finger2009robust} keeps the number of reads within a constant factor of the optimal.
Overall, the focus and the main challenge in  top-$k$ processing is ordering the answers according to their ranking scores ~\cite{IlyasBS08}. In the unranked case, when this challenge is absent, the rich body of work on top-$k$ processing does not go beyond the initial observations made by Carey and Kossmann~\cite{Carey1997}. 
NeedleTail \cite{Kim2018,Kim2018tech} specifically focuses on providing \emph{any-$k$} answers to queries without \texttt{ORDER BY} clauses, but it only handles key-foreign key joins, which dominate in the OLAP scenarios. In contrast, we support arbitrary CQs (i.e., select-project-join queries), allowing the complexity to grow with the tree-width (Theorem~\ref{thm:answers_upto}). Moreover, any-$k$ evaluation of CQs is just a building block of the processing of much more general threshold queries.

\newstuff{
\paragraph{Runtime optimization}
A large body of research on query processing led to powerful optimization techniques, such as \emph{aggregate pushing}~\cite{YanL95,ChaudhuriS94,ChaudhuriS96,genproj} and \emph{sideways information passing}~\cite{BernsteinC81,ChenHY97,IvesT08,MackertL86,OrrKC19}. These techniques aim to speed up the execution of a given join plan and rely on a cost model to heuristically approximate instance-optimal plans. 
Our focus is on reducing the search space of the heuristic methods by identifying plans with good worst-case guarantees. Such plans can be further improved towards instance-optimality, using classical techniques. For \limit queries the combination of sideways information passing and \limit pushing might be beneficial. Indeed, if we can ensure that each tuple produced by the subplan extends to a full answer, then we can stop the execution of the subplan when the desired number of tuples is output. For general threshold queries, the potential for such optimization is less clear. There, instead of a global limit on the number of answers we have a per-group limit. Consequently, the execution of the subplan can be stopped only when each group has sufficiently many tuples. The level of savings depends on the order in which the subplan produces its results. 
Such optimization goes beyond the scope of our paper, but is a promising direction for future work.
}

\paragraph{Quantified Graph Patterns}
Fan et al. \cite{FanWX16} introduced \emph{quantified graph patterns} (QGP) that allow expressing nested counting properties like having at least 5 friends, each with at least 2 pets. In contrast to threshold queries, QGPs are unable to count $k$-hop neighbours for $k\geq 2$, nor can they count tuples of variables. Moreover, QGPs adopt isomorphism matching, while threshold queries follow the standard semantics of database queries. 

\paragraph{Aggregate Queries}
In the context of factorized databases, Bakibayev et al. \cite{bakibayev2013Aggregation} observed that pushing aggregation down through joins can speed up evaluating queries. These results can be reinterpreted in the context of tree decompositions \cite{OlteanuZ15}, but they optimize different aggregates in isolation and do not investigate the interplay between counting and existential quantification.
AJAR (aggregations and joins over annotated relations) \cite{joglekar2016ajar} and FAQs (functional aggregate queries) \cite{KhamisNR16} are two very general sister formalisms capturing, among others, CQs enriched with multiple aggregate functions. Because different aggregate functions are never commutative, the evaluation algorithms for both these formalisms require decompositions consistent with the order of aggregating operations. For example, when applied to counting answers to CQs, this amounts to free-connex decompositions, as in our Proposition \ref{prop:counting-freeconnex}. 
In contrast, we remove the free-connex assumption by showing that counting up to a threshold and existential quantification can be reordered at the cost of keeping additional information of limited size. 

\paragraph{Counting Answers}
For \emph{projection-free} CQs, the complexity of counting answers is tightly bound to tree-width \cite{DalmauJ-tcs04,FlumG04}, just like in the case of Boolean evaluation ~\cite{GroheSS-stoc01,Grohe-jacm07}, but the presence of projection makes counting answers intractable even for \emph{acyclic} CQs~\cite{Pichler_Skritek_2013}. Efficient algorithms for counting answers require not only low tree-width but also low star-size  \cite{DurandM15}. However, when the problem is relaxed to randomized approximate counting, low tree-width is enough \cite{arenas2020approximate}, just like for Boolean evaluation.  Our results imply that for a  different relaxation --- counting exactly, but only up to a given threshold --- CQs of low tree-width can also be processed efficiently. However, we go far beyond CQs and show how to count answers to threshold queries (which themselves generalize counting answers to CQs up to a threshold). 

\paragraph{Enumerating Answers}
Also in the context of constant-delay enumeration low tree-width is not enough to guarantee efficient algorithms: the query needs to have low \emph{free-connex tree-width} \cite{BaganDG07}. Importantly, even acyclic queries can have very large free-connex tree-width.
Tree-width with can be replaced with fractional hypertree-width \cite{OlteanuZ15,DeepK18,KaraO18} or  submodular width \cite{BerkholzS19} but always in the restrictive free-connex variant.
Tziavelis et al.~\cite{TziavelisAGRY-pvldb20,tziavelis2020TODS} partially lift these results to the setting of \emph{ranked enumeration}, where query answers must be enumerated according to a predefined order; the lifted results allow enumeration with \emph{logarithmic} delay and handle projection-free CQs of low submodular width as well as free-connex acyclic CQs (but not general CQs).
In this work, we show that if the number of needed answers is known beforehand, general CQs of low tree-width can be processed efficiently even if they have large free-connex tree width. Moreover, this result is only the starting point for processing general threshold queries, for which we also provide constant-delay enumeration algorithms. 

\paragraph{Sampling Answers}
Sampling query answers was identified as an important data management task by Chaudhuri at al. \cite{ChaudhuriMN99}, who proposed a simple algorithm for sampling the join $S \bowtie T$ by sampling a tuple $s \in S$ with weight $|T \ltimes \{s\}|$ and then uniformly sampling a tuple $t \in T \ltimes \{s\}$. Using the \emph{alias method} for weighted sampling \cite{Vose91,Walker77}, this algorithm can be implemented in such a way that after a linear preprocessing phase, independent samples can be obtained in constant time. 
This approach was generalized to acyclic projection-free CQs \cite{ZhaoC0HY18}. We extend the latter result in three ways: we handle non-acyclic CQs, allowing the complexity to grow with the tree-width; we can allow projection, at the cost of replacing tree-width with its faster growing free-connex variant; and we handle threshold queries, rather than just CQs. 
A different approach to non-acyclic projection-free CQs  \cite{Chen20} provides a uniform sampling algorithm with guarantees on the \emph{expected} running time; 
this is incomparable to constant-time sampling after polynomial-time preprocessing, offered by our approach. Finally, Arenas et al. \cite{arenas2020approximate} show that efficient \emph{almost uniform} sampling is possible for CQs of low tree-width. Here, \emph{almost uniform} means that the algorithm approximates the uniform distribution up to a multiplicative error; this is a weaker notion than uniform sampling. Let us also reiterate that the all these papers only consider CQs, not threshold queries. 


\section{Conclusions and Future Work}
\label{sec:conclusions}

In this paper, we have embarked on a deep theoretical study of a newly identified class of \emph{threshold queries}. Our extensive empirical study shows that threshold queries are highly relevant in practice as witnessed by their utility in real-world knowledge graphs and their presence in massive query logs. Our theoretical investigation shows that threshold queries occupy a distinctive spot in the landscape of database querying problems. Indeed, our complexity analysis proves that threshold queries allow for a more efficient evaluation than solutions for closely related problems of counting query answers, constant-delay query answer enumeration, and top-$k$ querying. 

As one of the first future steps, we intend to gauge thoroughly the performance of the proposed algorithms. Designing adequate protocols for experimental evaluation requires a deep understanding of the relationships between evaluation of threshold queries and the related querying problems, which we have already accomplished in the present paper.  We intend to carry out a comprehensive implementation of threshold queries in an existing database system similarly to how algorithms of top-k queries have been implemented and evaluated~\cite{LiCIS05,LiCIS05}. More precisely, we will implement dedicated threshold-aware variants of relational operators and then we will introduce them in the query planning stage. 

Our work has led us to identify remarkable similarities in the query answering methods tackling counting and enumerating answers: they rely on various techniques for compiling out existentially quantified variables. We plan to pursue this discovery further and give full treatment to the emerging question: is there a unifying framework for assessing threshold queries and the related problems? Further theoretical results about threshold queries can be envisioned, such as establishing a dichotomy of evaluation complexity and identifying a condition under which evaluation is strongly polynomial, rather than pseudopolynomial. 




\begin{acks}
This work stems from a larger effort on developing schemas for property 
graph databases, including cardinality constraints, led by the LDBC Property Graph Schema Working Group. We  thank Andrea Cal\`{i}, 
Leonid Libkin,
Victor Lee, 
Juan Sequeda, 
Bryon Jacob, and 
Michael Schmidt
for their useful comments and early feedback.
\end{acks}


\bibliographystyle{ACM-Reference-Format}
\bibliography{main.bbl}

\onecolumn

\end{document}